\documentclass[12pt]{article}
\usepackage{amsmath, amsfonts, graphicx, graphics, bm, float, adjustbox, lscape, threeparttable, rotating, makecell, setspace, tabularx, multirow, booktabs, natbib}

\usepackage[utf8]{inputenc}
\usepackage[top=3cm,left=3cm,right=3cm,bottom=3cm]{geometry}
\usepackage[graphicx]{realboxes}
\usepackage [english]{babel}
\usepackage [autostyle, english = american]{csquotes}
\usepackage{subcaption}

\MakeOuterQuote{"}
\usepackage[toc,page]{appendix}

\usepackage[T1]{fontenc}

\usepackage{hyperref}
\hypersetup{
colorlinks=true,
citecolor=magenta,
linkcolor=magenta,
filecolor=magenta,      
urlcolor=magenta,
}

\usepackage{lipsum}

\newcommand\blfootnote[1]{%
  \begingroup
  \renewcommand\thefootnote{}\footnote{#1}%
  \addtocounter{footnote}{-1}%
  \endgroup
}

\title{What do Firms Gain from Patenting? \\
\Large The Case of the Global ICT Industry.}

\author{
  Dimitrios Exadaktylos \footnote{\href{mail to://d.exadaktylos@imtlucca.it}{d.exadaktylos@imtlucca.it}. Laboratory for the Analysis of Complex Economic Systems, piazza San Francesco 19 - 55100 Lucca}  \and Mahdi Ghodsi \footnote{\href{Mail to:// ghodsi@wiiw.ac.at}{ghodsi@wiiw.ac.at}. Vienna University of Economics and Business (WU); the Vienna Institute for International Economic Studies (wiiw) www.wiiw.ac.at, Rahlgasse 3, 1060 Vienna, Austria.}
  \and Armando Rungi \footnote{\href{mail to://armando.rungi@imtlucca.it}{armando.rungi@imtlucca.it}. \textbf{Corresponding author.} Laboratory for the Analysis of Complex Economic Systems, piazza San Francesco 19 - 55100 Lucca} \blfootnote{\textbf{Acknowledgement}: Authors acknowledge funding by the Anniversary Fund of the Oesterreichische Nationalbank (Project No. 18128). Support provided by Oesterreichische Nationalbank for this research is gratefully acknowledged.}
  }

\date{This version: July 2024}

\begin{document}

\maketitle
\vspace{0.01in}
\begin{abstract}

This study investigates the causal relationship between patent grants and firms' dynamics in the global Information and Communication Technology (ICT) industry, as the latter is a peculiar sector of modern economies, often under the lens of antitrust authorities. We exploit matched financial accounts and patent grants in 2009-2017 by 179,660 companies in 39 countries. Preliminarily, we find that less than 2\% of larger firms are responsible for 89\% of the grants. We propose a quasi-experimental strategy that first controls for reverse causality and then separates the impact of IPR protection from the innovative content of inventions making use of exogenous variation at the patent offices. We find that patents have a considerable impact on market shares and the size of smaller companies (31.7\% and 30.7\%, respectively) in the first year after the grants, which is mainly due to IPR protection. Most of the bigger firms' gains fade away after controlling for reverse causality and endogeneity. Notably, we never observe a direct impact on profitability for any firm size category. Eventually, we argue that IPR reform proposals should consider firms' heterogeneity and improve IPR access for smaller companies to enhance competition.

\footnotesize
\noindent  
\\
\noindent\textbf{JEL Classification}: O31, O34, L22, L25, F23\\
\noindent\textbf{Keywords}: intellectual property rights, ICT, market competition, SME \\
\bigskip
\end{abstract}
\setcounter{page}{0}
\thispagestyle{empty}

\onehalfspace
\section{Introduction}
\label{sec: introduction}

Over the past decades, digitalization has played a significant role in the transformation of many production processes. Companies operating in the Information and Communication Technologies (ICT) industry have become major global players, while the digital sector has rapidly grown. The industry contributes innovative consumer goods and services and technological inputs for firms across many other sectors. The benefits of investing in ICT are evident because many firms can potentially gain in terms of efficiency \citep{brynjolfsson2003computing} through reshaping innovation strategies \citep{nambisan2019digital}. Thus, policymakers tend to attribute a high value to the ICT global industry as an engine of economic growth. However, concerns about a fast market concentration among a few Big-Tech global players have been raised. Thus, antitrust authorities in the US and the European Union started their probes to check for abuses. Regarding mechanisms on how market concentration is built and preserved, the regime of Intellectual Property Rights (IPR) is among the prime suspects, as critics assume that companies obtain unfair market advantages from excessive protection granted through patents.

Against the previous background, we aim to empirically investigate the impact of patenting activity on firms' dynamics in the global ICT sector in 2009-2017. We adopt a quasi-experimental design to focus on firms' market shares and understand market advantages from a patent grant, while other firms' outcomes (turnover, employment, productivity, profitability, and capital intensity) help identify the sources of that market advantage. For our purpose, we exploit a sample of 179,660 firms active in 39 countries in 2009-2017 with matched information on financial accounts and patent grants. First, we provide insights into ICT firms' heterogeneous patenting activity over time and across geography. Notably, we show that IPRs are highly concentrated in a few portfolios since less than $2\%$ of larger firms are responsible for about $89\%$ of patent grants. Moreover, most active patentees are in the United States and Asia, while EU firms lag in patenting ICT. When we drill down into the details, we find that very large firms in the US with more than 1,000 employees register up to 213 new patents on average in our period of analyses. Therefore, when we look at financial accounts, we find that the number of patent grants positively correlates with market shares, productivity, firm size and capital intensity.\\

Motivated by preliminary evidence, we challenge correlations by adopting a composite empirical strategy. Our main interest is: 
\begin{enumerate}
    \item to understand the direction of causality, i.e., whether ICT firms grow on the market thanks to IPR or whether they obtain patent grants because they are already big and productive;
    \item to separate the impact of IPR \textit{per se} on firms' dynamics from the effect induced by the innovative content of the registered inventions.
\end{enumerate}

The first is a classical problem of reverse causality, for which we apply a diff-in-diff strategy for panel data settings recently proposed by \cite{Callaway_Santanna_2021}, which is useful to us because we have multiple treatment periods and variation in treatment timing. The method also allows controlling that the assumption of parallel trends is valid, conditional on observed companies' characteristics. Thus, having complete information on the timing of patents' registration processes, we consider a firm treated if it has been granted patents in our period of analyses. 

Interestingly, stronger correlations between firm-level outcomes and patenting activity fade away after we challenge reverse causality. We find that the impact of patent grants is higher and more significant in the case of smaller firms, i.e., after considering firms with a number of employees below the median value. Although there are not many smaller firms that come to get a patent grant, once they get one, their average market shares ($31.5\%$), sales ($33.7\%)$ and employment ($30.8\%$) increase. Notably, in the case of larger firms, we record a lower average impact on market shares ($5.2\%$), sales ($5.7\%$), and employment ($4.5\%$). 

In the second stage of our analyses, we are interested in disentangling the impact of IPR protection. We presume that previous diff-in-diff results confound the latter with the effects of the innovation content of the inventions protected by the patents. Companies could have increased market shares simply by selling more innovative products and not necessarily because they obtained IPR protection. From another perspective, we argue that the ability to ask for a patent and obtain a grant from a patent office is also correlated with the ability to generate innovative products. Thus, pairwise correlations with market shares and firm size may be spurious. Therefore, to distil the impact of IPR protection, we propose a novel instrumental variable strategy based on information collected at the level of patent offices. Our strategy exploits the exogenous experience of non-ICT firms in getting grants in the same ICT technology-office-year cells as ICT firms. In fact, we find that it is quite common for companies to obtain patents in technologies that are not strictly related to their core activities. In our case, it is possible for a non-ICT firm to ask for a patent in a technology related to the ICT industry, thus undergoing the same evaluation process by experts at the patent offices. Our intuition is that companies in other sectors do not compete with those in the ICT industry, hence our exclusion restriction assumption for the validity of our instruments. Yet, regardless of a company's industrial affiliation, experts shall evaluate with the same criteria whether the inventions are innovative and thus worthy of protection. That's why we expect the propensity of ICT firms to obtain a grant in a specific technology in a given year to be correlated with the number of grants and/or the share of grants out of total applications in the same technology and the same year (relevance assumption). Eventually, both tests for the exclusion restrictions and the relevance assumptions confirm the validity of the chosen instruments.\\

After implementing our IV strategy, we find that the positive impact on a firm's market share and size is still there for smaller companies. In comparison, bigger companies do not show any significant impact after controlling for endogeneity. Notably, an increase in market share after a patent grant in a small company is relatively high (29.7\%) and statistically significant in the first years after the grant. Thus, we conclude that most of the impact of patent grants on smaller businesses can be \textit{ceteris paribus} attributed to IPR protection.

Finally, we argue that previous results underline the importance of considering firms' heterogeneity in the IPR system. Smaller firms show a positive and larger impact on their economic activity as they increase both sales and market shares. Yet, we have seen that only a few of the smaller firms accessed IPR protection. The problem is, among others, recognised by \citet{SMEscoreboard}, which surveyed the Small and Medium Enterprises (SMEs) that had innovations to protect but did not access IPR protection because. They report that the main reasons for avoiding the patent offices are: i) a lack of knowledge of the IPR system, ii) the complexity and the cost of registration procedures, and iii) the complexity and the cost of court procedures in the case of IPR infringements.

We believe our results should inform IPR reform proposals because they challenge the idea that all companies benefit from patenting in the same way. On the contrary, policymakers should promote IPR protection for the sake of smaller firms, which have demonstrated to have problems navigating the complexities of the IPR system \citep{SMEscoreboard}. Bigger firms are, indeed, likely to use patenting as an entry market barrier for smaller competitors, i.e., to exclude the latter from introducing productive innovations. Bigger firms can also use patents to license out technologies and obtain royalties, perhaps to those same smaller firms that have difficulties in having access to the patenting system. The problem becomes more relevant after the emergence of patent thickets that are used to fragment unique inventions across different patents to maximize the rents from licensing. Clearly, smaller firms can find it more difficult to understand whether their inventions are novel when they have to navigate the complex contours of patent thickets. Eventually, we conclude that an IPR reform that empowers smaller firms will inevitably have pro-competitive effects for the entire ICT market.

Quite surprisingly, neither profits nor productivity seem to be affected by patent grants after controlling for reverse causality. Thus, we discuss the peculiarities of the ICT industry that arguably represents a poster child of a sector with endogenous sunk costs\footnote{For a useful review on the different approaches to investigate market power in modern economies with an industrial theoretical perspective, see \citet{Berry_etal_2019}.}. Following the seminal works by \citet{Sutton1998}, we argue that ICT global players have the incentive to invest in R\&D to increase market shares since consumers evaluate product differentiation by looking at innovative features available on the market. Yet, given high investments in R\&D, successful patent grants do not allow assignees an immediate increase in capital returns \textit{per se}. On the contrary, most active patentees in the ICT global industry seem to show relatively more short-term financial distress in our data than smaller ones, possibly due to a rush in innovation.\\

Importantly, we make our findings robust to different corporate control strategies since we are aware that ICT headquarters can often delegate subsidiaries to hold their IPRs. It is the case when R\&D labs are located within corporate boundaries, for example, in the case of multinational enterprises that use their subsidiaries across national borders to invest in innovation and to hold patent portfolios.

To grasp the relevance of patenting activity in the ICT global industry, we report in Table \ref{tab: top_firms} a match of the top ICT global firms according to Fortune Global 500 in the reference year 2020 with the stock of patents they have accumulated over time, as from our matched data set\footnote{Please note that in following analyses we will always focus on flows of grants by year, while Table \ref{tab: top_firms} reports stocks of grants accumulated up to 2020, i.e., including grants that have been obtained since the incorporation.}. The Fortune's ranking is originally based on global revenues, and consistently, we match in the last column with information on all the patents that could have been obtained historically by either a parent company in the origin country or its subsidiaries located wherever in the rest of the world.

\begin{table}[H]
\caption{Top ICT global firms and stocks of patents}
\centering
\resizebox{0.75\columnwidth}{!}{
\begin{tabular}{clcccc}
\hline \hline \\
         Fortune's 500 & Company & Country & Revenues & N. employees & N. granted\\
         Global rank &  &  & (bln USD) &  & patents\\
\hline \\
\vspace{0.1in}
1 & Apple & United States & 260,174 & 137,000 & 54,536\\\vspace{0.1in}
2 & Samsung Electronics & South Korea  & 197,705 & 287,439 & 641,743\\\vspace{0.1in}
3 & Foxconn & Taiwan & 178,860 & 757,404 & 2,266\\\vspace{0.1in}
4 & Alphabet & United States & 161,857 & 118,899 & 60,049\\\vspace{0.1in}
5 & Microsoft & United States & 125,843 & 144,000 & 89,635\\\vspace{0.1in}
6 & Huawei & China & 124,316 & 194,000 & 98,880\\\vspace{0.1in}
7 & Dell Technologies & United States &	92,154 & 165,000 & 11,509\\\vspace{0.1in}
8 & Hitachi & Japan & 80,639 & 301,056 & 268,598\\\vspace{0.1in}
9 & IBM & United States & 77,147 & 383,056 & 216,837\\\vspace{0.1in}
10 & Sony & Japan &	75,972 & 111,700 & 219,092\\\vspace{0.1in}
11 & Intel & United States & 71,965 & 110,800 & 91,214\\\vspace{0.1in}
12 & Facebook & United States &	70,697 & 44,942 & 12,381\\\vspace{0.1in}
13 & Panasonic & Japan & 68,897 & 259,835 & 384,817\\\vspace{0.1in}
14 & HP Inc. & United States & 58,756 & 44,942 & 61,715\\\vspace{0.1in}
15 & Tencent & China & 54,613 & 62,885 & 18,552\\\vspace{0.1in}
16 & LG Electronics & South Korea & 53,464 & 74,000 & 315,038\\\vspace{0.1in}
17 & Cisco & United States & 51,904 & 75,900 & 17,997\\\vspace{0.1in}
18 & Lenovo & China & 50,716 & 63,000 & 27,716\\
\hline \hline                
\end{tabular}%
}
 \begin{tablenotes}
      \footnotesize
      \item Note: The table indicates the list of top ICT global firms in the year 2020 according to the Fortune Global 500 ranking and the total number of patents that have been granted at any time in their business history in the same year, as reported by the Orbis Intellectual Property database.
   \end{tablenotes}
\label{tab: top_firms}
\end{table}

Notably, we record an average stock of about 160 thousand by top ICT firms; the most historically active assignee has been Samsung Electronics in South Korea, with up to 641,743 grants. The younger Foxconn in Taiwan is the one relying relatively less on patenting activity with an albeit non-negligible stock of 2,266 patents. In the following analysis, we will go beyond bigger corporations to investigate the role of medium and smaller companies while showing how patenting activity can be heterogeneous within the global industry in a relationship with firm-level outcomes.
\\

The rest of the paper is organized as follows. Section \ref{sec: literature} relates our work to previous literature. Section \ref{sec: data} introduces data and provides preliminary evidence extracted from our matched patent-firm sample. In section \ref{sec: results}, we discuss our identification strategy and comment results. Section \ref{sec: robustness} introduces some robustness and sensitivity checks. Finally, Section \ref{sec: conclusion} concludes with final remarks.

\section{Related literature}
\label{sec: literature}

First of all, we relate our contribution to the vast strand of literature that investigates the impact of IPR protection. The latter is usually justified as a way to introduce artificial scarcity and amend non-rivalry and non-excludability in the consumption of knowledge. The economic rationale of IPR protection is that early positive externalities reduce the incentives for knowledge producers who may underinvest in an industry that significantly contributes to social welfare and economic growth. In this context, patents are supposed to be a way to counterbalance market imperfections while generating a temporary legally enforced monopoly to guarantee producers profits from knowledge generation. 

Over the last years, however, several scholars have raised concerns about the perverse mechanisms that IPR practices can bring. In their seminal works, \cite{dosi2006much} and \cite{boldrin2008against} build cases against intellectual monopolies, discussing evidence that IPR regimes have, at best, no impact or, in some cases, a negative impact on innovation rates. IPRs may favour rent-seeking behaviour by firms that benefit from the monopolistic power granted on the knowledge they generate, therefore reducing the diffusion of innovation and reducing social welfare. Interestingly, for our case, \cite{Boldrin&Levine2013} point out how there seems to be no positive relationship between patenting activity and productivity. Expressly, the authors point to an inconsistency between the partial equilibrium, where patents may still be able to raise incentives for incumbent producers, and the general equilibrium, where protection can reduce aggregate innovation rates.

Looking into specific domains, \cite{henry2010intellectual} discuss solutions for slowing down climate change and enhancing environmental protection, where patenting is most problematic. In particular, they sketch the case study of research in genetically modified organisms, where a different regime has brought more comprehensive social benefits. Interestingly, \cite{moser2013patents} and \cite{van2018patent} review other cases in economic history when, in the absence of modern IPR regimes, different forms of protection or knowledge sharing could also accompany waves of essential innovations. Eventually, \cite{cimoli2014intellectual} discuss how modern IPR regimes could represent an obstacle to knowledge diffusion in developing countries, which may need to imitate prosperous developed countries to boost economic growth. 

Yet, on the other front of the controversy, there are scholars that stress how IPR protection is even more critical in modern times if one considers the strategic role that intangible assets play for the economic potential of regions and countries \citep{ziedonis2008apparent, haskel2018capitalism}. Against the previous background, we argue that today's IPR protection is still needed. It rather needs to focus more on the more vulnerable subjects, e.g., smaller firms, which find it more costly to accede to patent offices and to pay at the court in the case of infringements \citep{SMEscoreboard}. In fact, we find that the benefits from patenting can be evident for smaller firms, which react as one would expect, i.e., by increasing firm size and share in the first year after the grants are obtained. Yet, only a small share of smaller firms actually obtain grants in the period of analyses, and bigger firms hold the biggest portfolios of grants. Ideally, we believe that any IPR policy reform should consider the role that firm heterogeneity has in the access to the IPR system is relevant.

Of course, the role of IPRs in the ICT industry has been the subject of many previous studies\footnote{In Appendix Table \ref{tab: lit review}, we summarize the main issues and authors that we think are relevant to our study}\footnote{For a comprehensive review, see \citet{Comino_et_al_2018}.}. For our scope, we only need to underline a few milestones about the evolving relationship between IPRs and ICT. The industry has been responsible for a patent explosion since the 1980s \citep{Hall_2004}. \citet{Danguy_et_al_2014} show that a patent explosion cannot be attributed to a surge in research productivity. Rather, it is the product of the globalization of IPRs, since companies can decide that it is convenient to claim their rights in front of different patent offices. Yet, when \citet{Venturini_et_al_2022} focus on so-called digital intelligent technologies, they noted how they contributed to productivity because they helped implement the Fourth Industrial Revolution (4IR). Thus, they estimate that the segment of intelligent digital technologies accounts for a range between 3\% and 8\% of observed aggregate productivity changes in a sample of industrialized economies in 1990-2014.

Notably, a surge of patents in ICT correlates with the fragmentation of IPRs. Patent thickets emerge because ICTs are complex technologies where innovation is cumulative, and improvements or recombinations of previous inventions are relevant. Thus, ICT assignees often have overlapping claims through patent thickets \citep{Graevenitz_et_al_2013}. More than often, property rights on different technological system components are dispersed among several operators, and single companies may want to secure all the licenses needed to \textit{'hack their way through the patent thicket'} \citep{Shapiro_2001}. Yet, from our perspective, patent thickets further complicate the life of smaller firms, which may find it more difficult to understand how their inventions relate to the fragmented patents that are already registered. Although several entities (patent pools, standard-setting organizations, and patent intermediaries) have emerged that help overcome patent fragmentation, especially when it is important to establish technological standards \citep{Blind_et_al_2023}, we know that smaller firms still perceive the IPR system too difficult to navigate \citep{SMEscoreboard}. 

In fact, bigger ICT companies may use patents less to protect focal innovations and more as strategic tools in negotiations, litigations, and signalling their technological position. From a legal perspective, we can say that, in many situations, patents have become a way to preempt substitute inventions \citep{Cappelli_et_al_2023} and license them out in a consolidated market for ideas \citep{Gambardella_et_al_2007, Arora_Gambardella_2010}. In this context, it makes sense that we find correlations between firm size and patent portfolios, which, however, fade away when we control for reverse causality because additional patents are not used to record a direct impact on bigger ICT producers' outcomes. Not surprisingly, producers seem particularly keen on claiming IPR protection through court proceedings \citep{graham2013smart} in order to preserve their market advantages, and the cost of going to court can be relatively higher for smaller patentees.

Notably, we also refer to recent literature that underlines how peculiar is the case of the ICT global industry. The ICT industry contributes to global economic growth \citep{Nguyen_Doytch_2022}, thanks to the widespread adoption of technologies that enhance the productivity of both private and public activities \footnote{Please note, however, the existence of a strand of research that questions the actual contribution of modern ICTs to aggregate productivity as unsatisfactory if measured against initial expectations. The argument follows that one should expect much more productivity from adopting new technologies than what is measured, hence a so-called productivity paradox. Among others, see \cite{Acemogluetal2014}.}. Firms in ICT have unique business models and require technological platforms that engage many downstream producers \citep{teece2018profiting}. Given the relevance of innovations coming from the ICT industry, bigger producers seem particularly keen on claiming IPR protection through court proceedings \citep{graham2013smart}, thus keeping their market advantages against smaller competitors.

Against the previous background, we also refer to the research strand based on the seminal works by \citet{Sutton_1986, Sutton1991, Sutton1998}, who support the idea that there are industries that operate with endogenous sunk costs, where there is a structural lower bound in market shares. Our results show no direct causal impact on profitability from patenting. We believe these results are largely expected in innovative industries like ICT, where firms have to invest a relatively higher share in R\&D, advertising and other activities to enhance consumers' willingness to pay for the products they offer, independent of how big the demand already is. See also previous ideas in \citet{Shaked_Sutton_1983, Shaked_Sutton_1987}. In this case, a market equilibrium implies a relatively higher market share that is associated with higher technological barriers to market entry and, thus, a smaller pool of firms that can profitably operate in the industry. Efforts by ICT companies to reach consumers with the most innovative products may not be rewarded \textit{per se} with increasing profits, although such innovative activities are necessary to outlive the ICT market.\\

For a competing framework, see also \cite{acs1988innovation} and \cite{geroskipomroy1990innovation}, who underline how innovation may be negatively associated with market concentration. \cite{aghion2005competition} suggest that firms have a market advantage when they innovate in industries that suffer from lower competition. Otherwise, when competition is higher, market followers have lower incentives to innovate than the leaders. On the same line, \cite{blundell1999market} also challenge the association between market share and innovation. After exploiting dynamic count data models, they found a robust and positive effect of market shares on patent stocks, although increased product market competition in the industry tends to stimulate innovative activity.

Beyond the ICT industry, we believe our results can relate to previous evidence on firms' outcomes from US industries published by \cite{balasubramanian2011happens}, according to which increases in patent stocks are associated with increases in firm size, scope, skill intensity, and capital intensity. Please note, however, how \cite{balasubramanian2011happens} did not test any impact on market shares, thus leaving the reader agnostic about the consequences of IPR on market structures.\\

More controversial is the relationship between patenting and productivity, which is never robust to reverse causality in our study. In the US, \cite{balasubramanian2011happens} find only a weak significance of the nexus after using data similar to ours matched at the firm-patent level. Unfortunately, we can only loosely relate to previous studies because we cannot retrieve similar indicators of R\&D expenses \citep{griffith2006innovation,  mairesse2009innovation, mohnen2013innovation, crespi2012innovation}. Eventually, we could recall the work by \cite{bloom2002patents}, who find that patents could have a significant impact on firm-level productivity only in the longer run, once inventions are incorporated into the production processes and efforts have been made to promote new products or production processes. Yet, from our viewpoint, the empirical evidence provided by \cite{bloom2002patents} is not entirely convincing. We argue that the authors test their hypotheses on a highly self-selected sample of only about $200$ firms that can stay quoted at the stock exchange throughout the entire period of analyses, thus not representative of the underlying business population that would include smaller and medium-sized firms.

In conclusion, please note that we always make our analyses robust to different definitions of the corporate perimeter, thus encompassing patents that are either granted to parent companies or their subsidiaries. In this way, we can control for optimal strategies by multinational enterprises that can, for example, locate part of their R\&D activities in countries where IPR regimes are more favourable or where taxation is relatively lower \cite{skeie2017innovation} on R\&D activities. It is the case of  IPR regimes where patent boxes are allowed; thus, revenues from granted patents are exempted from taxes to benefit from higher profits from international activities \citep{bosenberg2017rnd, alstadsaeter2018patent, davies2020patent}. In general, there is ample evidence that domestic and multinational enterprises in any sector can also take advantage of technology developed across different geographic regions, thus exploiting local subsidiaries for reverse knowledge transfer \citep{driffield2016reverse}. Therefore, an exclusive focus on parent companies would neglect an essential share of companies' innovation efforts.

\section{Data and preliminary evidence}
\label{sec: data}

\subsection{Data on firms and patents}

For our purpose, we exploit a matched data set of firms and registered patents in the period 2009-2017 sourced from the ORBIS database\footnote{The ORBIS database has become a standard source for global financial accounts. See for example \cite{gopinathetal2017}, \cite{CravinoLevchenko2017}, \cite{DelPrete&Rungi2017}, \cite{FattoriniGhodsiRungi2020}, and \cite{Exadaktylos_et_al_2024}. The coverage of smaller firms and details about financial accounts may vary among countries depending on the requirements of national business registries, as observed in \cite{kalemli2015construct}.}, compiled by the Bureau Van Dijk. In particular, the module on Intellectual Property links companies and other entities (i.e., assignees of IPR) to their original patent filings collected from PATSTAT, the global database maintained by the European Patent Office. Usefully for our scope, the IP module by Orbis follows: i) the evolution of each patent filing, from the publication to the moment the property right is granted; ii) the changes in property rights from one assignee to another, e.g., in case of companies' mergers and acquisitions. Previous users of the same database include \cite{noailly2015directing}, who study the effect of technological change on environmental performance, and \cite{alstadsaeter2018patent}, who investigate the determinants of patent registration. \cite{andrews2014resources} also use a similar matched patent-firm data set to identify the impact of first patenting on firm performance across industries and countries.

Although the IP module by Orbis includes patents and assignees from all over the world, we can keep only patents held by firms for which we have the basic financial information that we need for testing our hypotheses. Notably, we need to eliminate those countries for which we are not able to transform monetary values into real values because official statistics do not report macroeconomic deflators. Deflators are, thus, sourced either from national statistics offices, from Eurostat or from the OECD STAN database. We source deflators for gross output, intermediate inputs, and capital goods by country and sector of activity, respectively. In cases where deflators are unavailable at the two-digit or a more aggregate sector level, we use the GDP deflators at the country level. Deflators for Taiwan do not appear in the OECD or Eurostat, and we source them from the official local statistics office.

Eventually, we ended up with a sample of $179,660$ firms active in 39 countries and operating in the ICT industry. In Appendix Table \ref{tab: countries}, we enlist the countries covered from our sample. There is no official source against which we can easily validate our sample. We decided to check from international trade data, based on total exports of ICT goods and services recorded by \cite{OECD2024}, and found that sample countries cover about 86\% of the global trade in ICT.

For our scope, we define an ICT perimeter encompassing both firms and patents, as both come with different classification systems that do not always match. It is quite possible that ICT firms obtain patents in non-ICT technologies, as it is the case that non-ICT firms obtain grants in ICT technologies. Actually, the latter is a case that we want to exploit for the potential of an identification strategy in the following paragraphs. In the case of firms, we will consider as ICT firms those that belong to either manufacturing or services industries following the work made by \cite{benages}, who compiled the 2018 PREDICT database for the European Commission. In Appendix Tables \ref{tab: sectors}, we enlist NACE 2-digit industries included in our sample. In the case of patents and related applications to patent offices, we will consider ICT technologies based on a classification proposed by Eurostat and based on 4-digit subclasses of the International Patent Classification (IPC) system\footnote{For further details, please see the original document available at \url{https://ec.europa.eu/eurostat/cache/metadata/Annexes/pat_esms_an2.pdf}}.

\subsection{Preliminary evidence}

\begin{figure}[H]
\caption{Firm size in the ICT sector}
\centerline{\includegraphics[scale=0.6]{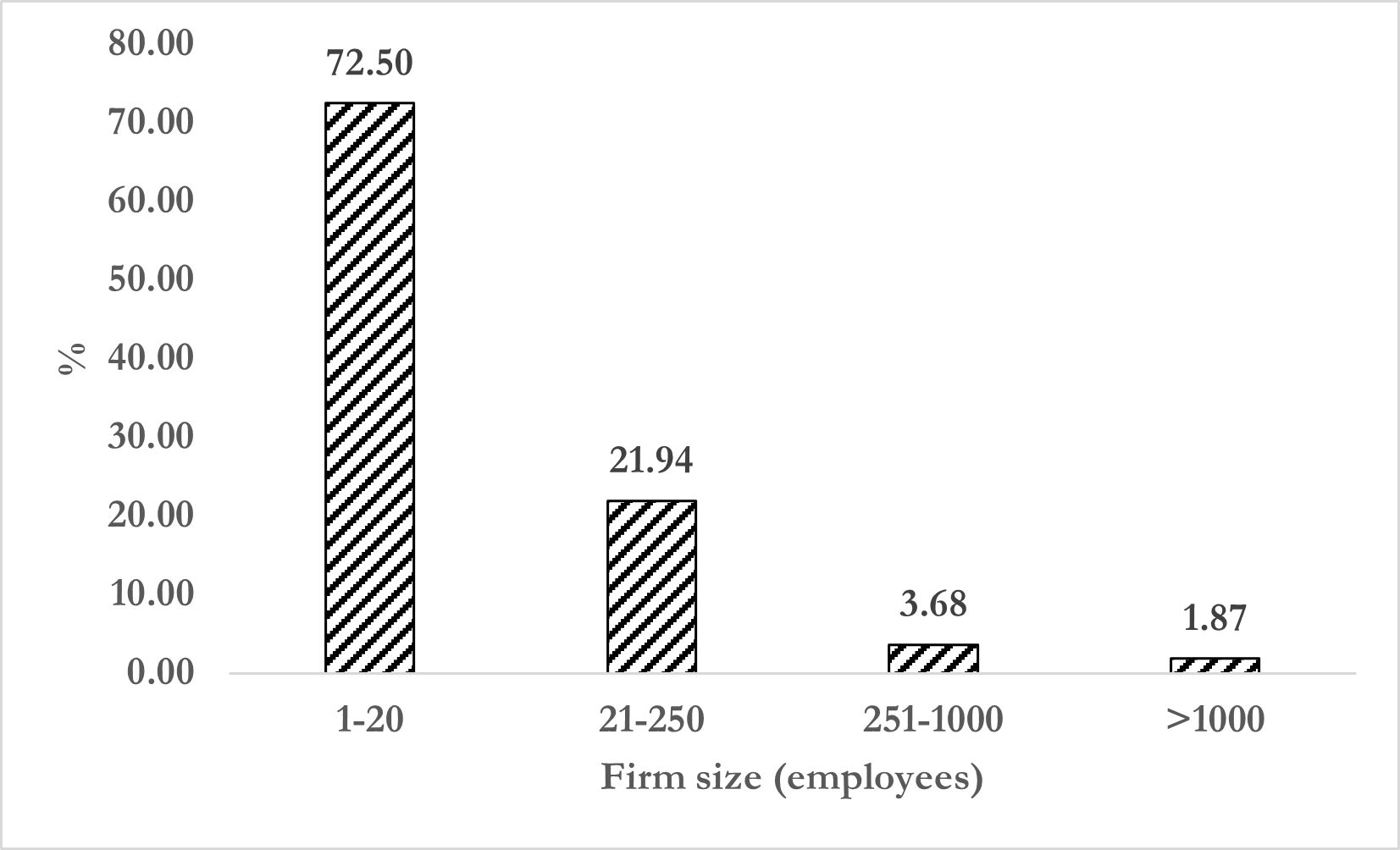}}
 \begin{tablenotes}
      \footnotesize
      \item Note: The figure shows the sample distribution of ICT firms by size (number of employees) in 2017.
   \end{tablenotes}
\label{fig: size_distribution}
\end{figure}

First, we provide a snapshot of ICT firms' heterogeneous distributions of patents, and we compare it with firm size distribution. Figure \ref{fig: size_distribution} shows the frequency of firms in four main size categories based on the number of employees in the last sample year: 1-20; 21-250; 251-1000; higher than 1,000. As expected, the largest category is the one represented by smaller firms ($74.1\%$), whereas bigger corporations with more than $1,000$ employees represent only $1.7\%$ of the sample. This aligns with most evidence about the heterogeneous distribution of firm size within any industry, not specifically of the ICT industry.

\begin{figure}[H]
\caption{Concentration of patent grants by firm size in the ICT sector}
\centerline{\includegraphics[scale=0.6]{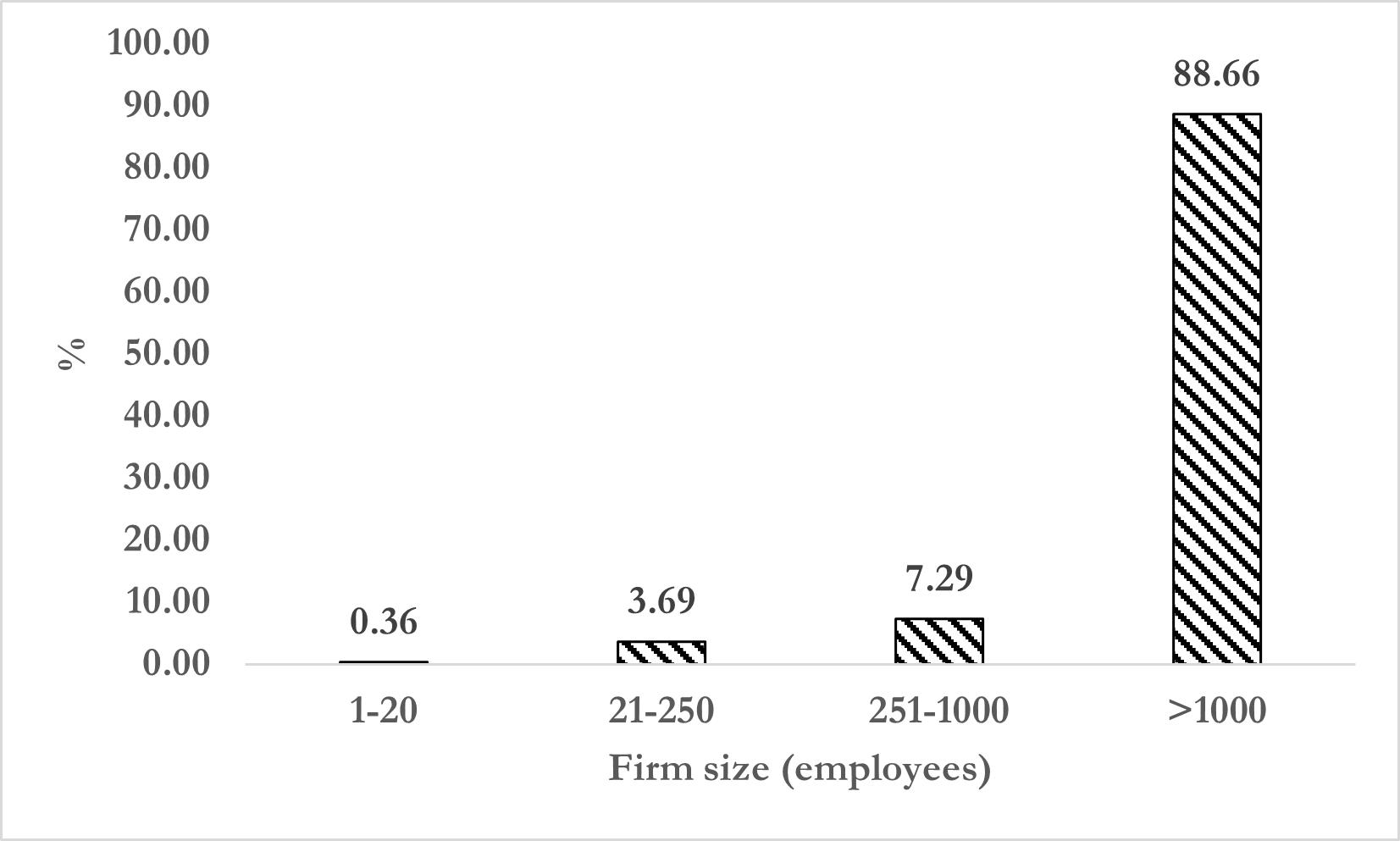}}
 \begin{tablenotes}
      \footnotesize
      \item Note: The figure shows the distribution of patent grants obtained by ICT firms in 2009-2017, in percentage by class of firm size. Firm size is measured as the number of employees in 2017. 
   \end{tablenotes}
\label{fig: patents_distribution}
\end{figure}

What is peculiar is the evidence reported in Figure \ref{fig: patents_distribution}, which shows that the category of very large firms, on the right of the distribution, actually accounts for $89.4\%$ of the patents that have been granted in our period of analyses. The latest is striking evidence that IPRs are highly concentrated in the ICT industry among a handful of larger players that can get a high number of grants, albeit in a short period of time. Notably, only a few firms active in the ICT industry are able to obtain patents, while a majority of them operate without getting any grants. We believe that such a concentration of IPRs among a few larger companies is preliminary evidence that is interesting \textit{per se}.\\

\begin{figure}[H]
\caption{Patents, firm size, and geography}
\centerline{\includegraphics[scale=0.75]{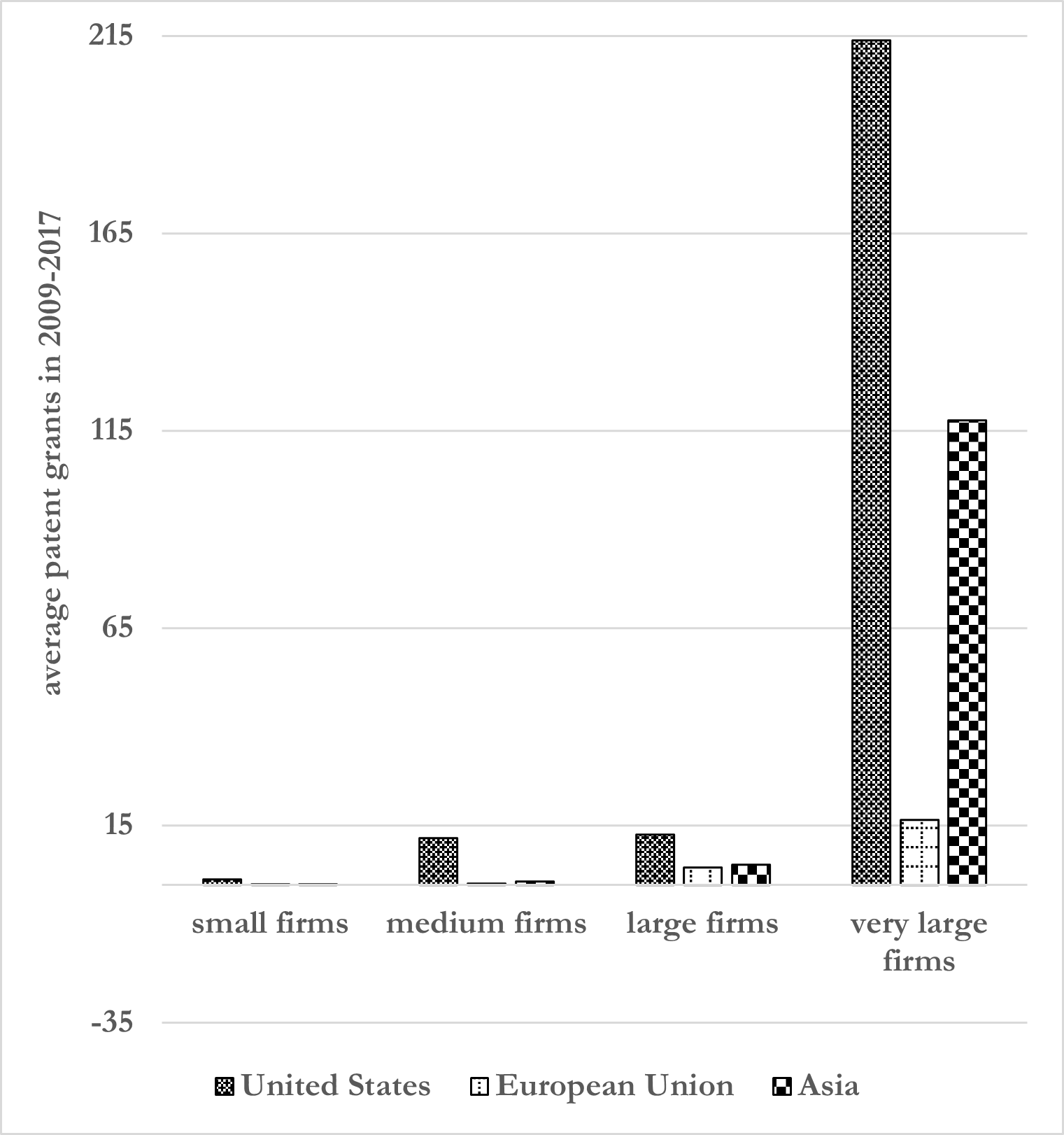}}
 \begin{tablenotes}
      \footnotesize
      \item Note: The figure shows the average number of granted patents by firm size and main geographic area in our period of analyses (2009-2017). Small firms are companies with up to 20 employees; medium firms with up to 250 employees; large firms with up to 1,000 employees; very large firms with more than 1,000 employees.
   \end{tablenotes}
\label{fig: patents geo size}
\end{figure}

Patenting activity is, at the same time, a sparse and concentrated phenomenon that involves mainly bigger companies in the ICT industry. Indeed, ICTs are complex technologies that require high R\&D sunk costs to operate. Such evidence motivates the following analyses to qualify how patent assignees look different from non-assignees, where they are, and the impact of patent grants on their firm-level outcomes. Let us start in Figure \ref{fig: patents geo size} to check where most active assignees are. After adopting the same classification by firm size of previous figures, we report on the x-axis the average number of grants that a firm obtained in 2009-2017. 

What we find is that the headquarters of the most active assignees are located first and foremost in the U.S., although very large firms with headquarters in Asia also obtain, on average, a relatively high number of patent grants, certainly higher than in the case of companies in the European Union.

Therefore, from our perspective, it is interesting to check whether and how assignees are statistically different from non-assignees. Table \ref{tab: ttest} provides preliminary evidence of firm-level outcomes' differences with t-tests. We observe that assignees have, on average, a much higher market share; they are significantly more productive, bigger and more capital-intensive than non-assignees. Yet, we do not observe a significant difference in profits as measured by ROCE (Return on Capital Employed)\footnote{Please note that we make our results robust to different definitions of profitability, both here and in the following paragraphs. Robustness checks on profitability are available upon request.}. The average assignee generates about $207,000$ dollars per worker in a year, whereas non-assignees register on average about $68,740$ dollars per worker in a year. The representative assignee generates about $7$ million dollars while average annual revenues are only $0.5$ million dollars when a company is not granted any patent.\\

\begin{table}[H]
\caption{Firm-level outcomes of patent assignees and non-assignees}
\centering
\resizebox{0.9\columnwidth}{!}{%
\begin{tabular}{lccccccc}
\hline \hline \\
 & \% Market share &  (log) Labour productivity  & (log) Size & (log) Capital intensity & ROCE (levels) \\
\\  \hline \\
Assignees & .0199 & 12.1624 & 17.6055 & 11.2350 & 0.0050 \\
          & (.0009) & (.0068) & (0.0158) & (0.0098) & (0.0023) \\
Non-assignees & .0004 & 11.1296 & 13.1669 & 8.8648 & 4.7349 \\
          & (.0001) & (.0014) & (0.0024) & (0.0023) & (4.6573) \\ \hline \\
Difference & .0195*** & 1.0329*** & 4.4386***  & 2.3701*** &  -4.7299 \\
          & (.0002) & (.0085) & (0.0147) & (0.0141) & (27.8492)\\\\  \hline \hline                
\end{tabular}%
}
 \begin{tablenotes}
      \footnotesize
      \item Note: The table reports t-tests on the differences in market share, (log) labour productivity, (log) size, (log) capital intensity and profitability (ROCE) for companies having at least one patent \textit{vis \'a vis} companies without patents. The unit of observation is firm-year level. Standard errors in parentheses. *** denotes significance at $1\%$.
   \end{tablenotes}
\label{tab: ttest}
\end{table}

Interestingly, in Appendix Tables \ref{tab: OLS parents' patents} and \ref{tab:OLS all patents corporate perimeter}, we also record that firm-level outcomes of interest positively correlate with the intensity of patenting activity. The higher the number of patent grants, the higher the market shares, productivity, firm size, and capital intensity. An exception is profitability measured by ROCE, which correlates negatively with patenting activity. Evidently, R\&D costs needed to innovate in the industry may strain financial accounts.

Clearly, empirical evidence so far does not say anything about the causality direction, i.e.,  whether it is the case that bigger and more productive firms are more able to obtain patents or whether patent grants allow them to gain market shares and become bigger and more productive. On the other hand, it is also possible that positive correlations so far are spurious, as both firms' outcomes and the ability to get patents correlate with the unobserved innovative content of the registered inventions. The following paragraphs will fundamentally challenge reverse causality and develop a strategy to unveil the endogenous relationships between firm-level outcomes, IPR protection, and innovative abilities.

\section{Empirical strategy and results}
\label{sec: results}

Our purpose is to assess the causal impact of patent grants on ICT firms' dynamics. In the previous paragraphs, we observed positive correlations between firm-level outcomes and patent grants. Our first step is to unravel the reverse causality that is possibly hidden in those correlations and say which comes first: whether it is the case that bigger and more productive firms are the ones that are more able to obtain grants or whether it's patent grants that allow them to gain market shares and become more productive. In this case, we will consider each firm that obtained a patent in a year as treated, as we would do in a quasi-experimental setting, to compare with a control group of firms that are not treated and, hence, investigate whether firm-level outcomes register an impact after the treatment. At this stage, we choose to adopt a difference-in-difference model introduced by \cite{Callaway_Santanna_2021}, which allows treatment to occur in multiple time periods. In Section \ref{sec: callaway}, it will be clear how this specific panel setup is ideal for our case, and we present results.\\

Then, in the second stage of our analysis, we aim to identify the contribution of IPR protection on firm-level outcomes, and separate it from the contribution of innovative inventions. It is indeed possible that observed positive correlations are spurious, as both firms' outcomes and the ability to get patents both correlate with the unobserved innovative content of the registered inventions. In this case, we cannot exclude that the innovative firm could have registered positive market outcomes regardless of patent registrations, thanks to the market potential of the registered inventions. Against this background, we introduce a novel instrumental variable (IV) approach, whose intuition is that we can exploit the exogenous variation that we can find on the set of non-ICT firms which have, however, registered ICT technologies in our period of analysis. In Section \ref{sec: IV}, we discuss the validity of our IV strategy.

\subsection{The impact of patent grants} \label{sec: callaway}

In this Section, we mainly draw on the econometric literature on treatment effects estimation \citep{Imbens_Woolridge_2009}, and we set up a quasi-experimental strategy. We consider a firm as treated if it has been granted a patent in our period of analysis. Thus, we can compare with a control group that is made of firms that were not granted a patent and, yet, they are as similar as possible to the treated firms. In this way, we can consider the outcomes of the control group as those of a counterfactual, which we could have observed if the patents were not granted to the treated firms. 

For our purpose, we implement a difference-in-difference on a panel data set following \cite{Callaway_Santanna_2021}. The aim is to identify the average treatment effect on the treated (ATT), i.e., the difference between the outcomes of the treated firms and the outcomes of the firms in the control group. The peculiarity of \cite{Callaway_Santanna_2021} is that we can obtain an ATT for any group of firms $g$ that obtained at least a grant at a specific time $t$, as follows:

\begin{equation}
    ATT(g,t) = \mathbb{E} \left[ \left( \frac{G_g}{\mathbb{E}{[G_g]}} - \frac{\frac{p_g(X)C}{1-p_g(X)}}{\mathbb{E}{\left[\frac{p_g(X)C}{1-p_g(X)}\right]}} \right) \left( Y_t - Y_{g-1} - m_{g,t}(X) \right)   \right]
\end{equation}

where $G_g$ is a binary variable equal to one if a firm belongs to the group $g$; $C$ is a binary equal to one for firms that have never been granted a patent at any time period; $Y_t$ is the firms' outcome at time $t$, i.e., market share, labour productivity, firm size, capital intensity or profits (ROCE). Then, $p_g(X) = P(G_g = 1|X, G_g + C = 1)$ is the probability of publishing a granted patent at time $g$ conditional on pre-treatment covariates $X$ and: i) either belonging in the group $g$; ii) or not being granted any patent at any time during the period. Then, $ m_{g,t}(X) = \mathbb{E}[Y_t - Y_{g-1} | X, C=1]$ is the population outcome regression for the control group made by firms that have never been granted a patent in our period of analyses. Please note that \cite{Callaway_Santanna_2021} provide alternative specifications to estimate group-time average treatment effects. In this application, we adopt the doubly robust estimator first proposed by \cite{santanna2020dr} because it is more challenging in identification than the alternatives. The matching procedure to derive the control group makes use of inverse probability weights after a propensity score is obtained with logistic regression. More precisely, the matching is done considering as covariates the capital intensity, the number of employees, firm age, 4-digit NACE rev. 2 industry-level dummies, and three location-specific fixed effects for headquarters located in the European Union, the United States, and the rest of the world. Importantly, we always check that the assumption of parallel trends is made conditional on companies' characteristics before the treatment occurs. The firms' characteristics that we pick for pre-treatment trends are the same that are used for the propensity score matching.

Please note that, for this exercise to work, we must consider a balanced panel with complete information on labour productivity, employment, capital intensity and age. Thus, we exclude companies that registered patents in 2009 because we cannot check for what happens before treatment in that year. Eventually, we end up with a reduced sample of $24,522$ firms, of which only $432$ companies have been treated at some point in 2010-2017, and $24,091$ have never been granted any patents in the same period\footnote{Please note that at this stage, we consider only patents granted directly to firms. In a robustness check reported in the Appendix Table \ref{tab: corporate perimeter shares}, we repeat the exercises by considering also treated those firms that were granted patents indirectly because they were held by one of their subsidiaries in the corporate perimeter. Results do not significantly change.}.

At this point, to estimate the overall impact of patenting on firm-level outcomes, we shall consider a weighted average of previously defined $ATT(g,t)$ in the following way:

\begin{equation}\label{eq: aggregate_ATT}
    \theta_s^O = \sum_{g=2}^T \theta_s (g) P(G=g)
\end{equation}

where,
\begin{equation*}
    \theta_s (g) = \frac{1}{T-g+1} \sum_{g=2}^T \mathbf{1}\{ g \leq t\} ATT(g,t)
\end{equation*}

and $T$ denotes the number of years. In other words, even if we work on a panel data set, where firms can be granted patents at different moments on the timeline, we can still obtain a unique parameter, $\theta_s^O$, which tells us whether patents have an impact on firm-level outcomes. That parameter is finally a weighted average of time-specific parameters, as the latter are obtained considering groups of firms that have been treated in any observed period. The group-specific weights, $P(G=g)$'s, are obtained considering the relevance of each group over the total sample. 

Finally, we can test the persistence of the effect thanks to a classical event study analysis, for which we need to compute the length of exposure to the treatment, $e$. The latter is another form of aggregation of the group-time specific effect, which we can define as:

\begin{equation} \label{eq: event study}
    \theta_{es} (e) = \sum_{g=2}^T \mathbf{1} \{g+e \leq T\}  P(G=g|G+e \leq T) ATT(g, g+e)
\end{equation}

In plain words, eq. \ref{eq: event study} returns the average impact on firm-level outcomes after $e$ periods from being granted a patent, considering the heterogeneity across all cohorts participating in the treatment.

In Table \ref{tab: baseline}, we report estimates of the impact of patenting activity on firm-level outcomes. In Appendix Table \ref{tab:balancing}, we also report a test that positively checks if continuous properties balance across the treatment and the control groups. According to our findings, companies being granted patents in the period 2010-2017 benefit from an increase in market share by $9.5\%$ (log units: $0.091$), which comes with a parallel increase in revenues ($9.7\%$; log units: $0.093$), as largely expected. Indeed, patents by ICT firms are mainly devoted to the protection of product innovations. Firms ask for the protection of new technological advancements that improve the products that they professionally sell. In this case, we argue that it makes sense that we detect an impact on market shares and firm size, thanks to higher revenues derived after IPR protection. We cannot exclude that firms in sectors different from ICT can derive a significant benefit from grants in productivity.

Yet, considering the heterogeneity in the distributions of both firm size and patenting activity observed before (Figures \ref{fig: size_distribution} and \ref{fig: patents_distribution}), it makes sense to check if the impact on smaller patentees is significantly different from the impact on bigger corporate players. In the second and third sections of Table \ref{tab: baseline}, we split the sample of firms considering the size (measured by revenues) before the treatment. We consider smaller firms the ones that report revenues below the sample median, and we separate them from bigger firms as they record revenues above the sample median. 

Results clearly show that the magnitude of the impact for smaller firms is quite relevant on both market shares ($31.5\%$; log units: $0.275$) and firm size measured as turnover ($33.7\%$; log units: $0.291$) and employment ($18.8\%$; log units: $0.172$). Notably, coefficients on labor productivity, capital intensity and profitability (ROCE) are never statistically significant. Evidently, after challenging reverse causality, a large part of the correlations of firm-level outcomes with patenting activity observed in Tables \ref{tab: ttest} and \ref{tab: OLS parents' patents} fades away. 

In other words, our findings show that productivity and profitability are not a direct consequence of the grants that firms obtained, at least in the short run. We observe that the latter findings are coherent with cross-country general evidence on manufacturing companies beyond the ICT industry, as in \cite{andrews2014resources}. On the other hand, our evidence contrasts with \cite{balasubramanian2011happens}, who find a significant effect on productivity in the case of U.S. firms, although relatively smaller if compared to the impact on firm size.

\begin{table}[H]
\caption{The Average Treatment Effect of patent grants on firm-level outcomes}\label{tab: baseline}
\centering
\singlespacing
\resizebox{0.8\columnwidth}{!}{%
\begin{tabular}{lcccc}
\hline \\
Variable & $\theta_s^O$ & s. e. &  N. treated firms & N.  untreated firms\\
\hline \hline \\
\textit{All firms} & & \\ \\
(log) Market share & 0.091*** & (0.031) & 432 & 24,090 \\ \\
(log) Labor productivity  & 0.003 & (0.023) & 432 & 24,090  \\ \\
(log) Firm size (employees)   & 0.067***  & (0.023) & 432 & 24,090   \\ \\
(log) Firm size (turnover) & 0.093*** & (0.031) & 432 & 24,090   \\  \\
(log) Capital intensity & 0.037 & (0.041) & 432 & 24,090  \\ \\
ROCE (levels) & -0.015 & (0.014) & 330 & 18,332   \\ \\
\hline
\\ \\
\textit{Small firms} & & \\ \\
(log) Market share & 0.275*** & (0.103) & 71 & 12,190 \\ \\
(log) Labor productivity & 0.059 & (0.079) & 71 & 12,190 \\ \\
(log) Firm size (employees) & 0.172*** & (0.059) & 71 & 12,190   \\  \\
(log) Firm size (turnover) & 0.291*** & (0.103) & 71 & 12,190   \\  \\
(log) Capital intensity & 0.117 & (0.120) & 71 & 12,190  \\ \\
ROCE (levels) & 0.010 & (0.053) & 50 & 9,570  \\ \\
\hline
\\ \\
\textit{Large firms} & & \\ \\
(log) Market share & 0.051* & (0.029) &  361 & 11,900 \\ \\
(log) Labor productivity  & -0.011  & (0.021)  &  361 & 11,900   \\ \\
(log) Firm size (employees)  & 0.044*  & (0.023) &  361 & 11,900  \\  \\
(log) Firm size (turnover) &  0.055* & (0.029) & 361 & 11,900   \\  \\
(log) Capital intensity & 0.015 & (0.042) &  361 & 11,900   \\ \\
ROCE (levels) & -0.009 & (0.016) & 280 & 9,570  \\ \\
\hline
\end{tabular}}
\begin{tablenotes}
\footnotesize
\item Note: The table illustrates average aggregate treatment effects after following the doubly robust version of the method by \citet{Callaway_Santanna_2021}, in the presence of a panel setting, under the assumption of parallel trends conditional on firm-level control variables, 4-digit NACE sector, and regional dummies. Treated firms are matched with untreated firms using inverse probability weights. Errors are clustered at the firm level. *, ** and *** denote significance at 10\%, 5\% and 1\% respectively.\\
\end{tablenotes}
\end{table}

Interestingly, we can see at the bottom of Table \ref{tab: baseline} that the impact on larger firms' accounts is statistically weakly relevant and with a significantly lower magnitude on market shares ($5.2\%$; log units: $0.051$), turnover ($5.7\%$; log units: $0.055$) and employment ($4.5\%$; log units: $0.044$). From our viewpoint, it makes sense that bigger ICT players benefit less from additional patent grants at the margin since they already have a relevant dimension and they already hold, on average, a consistent portfolio of older patent grants.

Moreover, please note that our findings do not exclude that companies also benefit indirectly from patenting activities, e.g., when using patents strategically in litigations, negotiations, and as a signal vs potential competitors. As already discussed in Section \ref{sec: literature}, the ICT industry is a typical sector with complex technologies where strategic registrations of patents emerged in the latest decades. In this context, indirect benefits can emerge, such as lower competition on product lines already under a grant. Potentially, indirect effects explain the strongest correlations between patenting activity and firms' outcomes, which we discussed in Section \ref{sec: data} before we challenge reverse causality. Of course, such indirect effects are, on purpose, eliminated from our difference-in-difference strategy because our aim is to focus on the direct effects of patenting. We will see in the next paragraphs that they do not keep their statistical importance after we focus on IPR protection with an IV strategy. In fact, previous coefficients may still confound the impact of IPR with the one played by innovation. Even if we controlled for the selection of most productive firms into the status of patentees, we could not separate the specific role of unobserved innovations on market outcomes. The main problem is that patent grants are both an indicator that companies were able to innovate and, simultaneously, of the legal protection they obtained from imitation by competitors that could otherwise challenge market shares.

Finally, we report event studies following the procedure described in eq. \ref{eq: event study} in Appendix Figures \ref{fig: MS_ES}-\ref{fig: ROCE_ES}. We aim to check how our main firm-level outcomes of interest evolve as time passes from when the representative company has been granted a patent. As in any classical event study, we align events on a reference period, $e=0$, which is the first year a firm has been granted a patent in our sample, $e=0$. Therefore, after following eq. \ref{eq: event study}, we plot the impact on the outcome of the representative company at any following period, thereby checking that previous trends are conditional on firm-level characteristics, industry affiliations, and the firm's location choices. Evidently, in any of the figures, we do not visualize any statistically significant trend before treatment, i.e., companies are not systematically showing that they were becoming bigger, more productive, or capital intensive before obtaining patent grants, in $e=0$.\\

\subsection{IPR protection and firms' dynamics: an IV approach}\label{sec: IV}

In previous paragraphs, we first detected a positive premium on firm-level outcomes by patentees over non-patentees. Then, we implemented a check for the self-selection of firms into the status of patentee with a difference-in-difference approach on a panel set. We concluded that smaller patentees gained an advantage in market shares and sales after the grant, and bigger patentees had, on average, smaller advantages. Still, we were agnostic about the economic channel that drives that impact. Gains in market shares and firm size are expected after a grant as either the result of a higher demand for innovative ICT products or as a consequence of the IPR protection from imitation by competitors. 

To focus on the role of IPR protection, we implement here a novel instrumental variable strategy. We propose two instruments to capture the innovative content of patent grants based on the information we have about assignees of ICT technologies that operate outside the ICT industry. Our intuition is that non-ICT firms do not compete with the ones in the ICT global industry; therefore, we do not expect firms' market shares and sizes to correlate across the two groups (exclusion restriction assumption). Yet, firms in non-ICT industries occasionally are interested in protecting their ICT innovations; therefore, when they apply for a grant, they are subject to the same evaluation process regardless of the assignees' economic activity\footnote{Please note how it is still possible that examiners evaluate more favourably the applications presented by ICT companies, better than non-ICT companies. The first may have more expertise in building a case for their inventions, or they can catch the attention of the examiners after lobbying for their interests. When we look into our data, we find that approval rates by ICT vs non-ICT companies are correlated about 82.7\%. Yet, t-tests show a 5.9\% higher probability that ICT companies get their approval on an ICT technology. In the context of our IV strategy, the difference in approval rates is acceptable as long as it does not make our instrumental variables weak. 82.7\% correlation already indicates a good albeit naïve test for the absence of weakness. More standard tests are reported in the regression tables}. (relevance assumption).

In particular, we consider: 
\begin{itemize}
    \item the approval rate of grants over the number of applications that non-ICT firms put forward in the same technologies of ICT firms, $j$, evaluated by the same patent offices, $h$, in each same year, $t$, as follows:

\begin{equation}
\textit{appr}_{jht}^{\ no\ ICT} = \frac{\textit{grants}_{jht}^{\textit{ no ICT}}}{\textit{appl}_{jht}^ {\textit{ no ICT}}} \nonumber \end{equation}

    \item the number of applications ($\textit{appl}_{jht}^{\textit{ no ICT}}$) by non-ICT firms at the same patent offices, $h$, in the same technology classes, $j$, and in the same year, $t$\footnote{Please note how from the application to the actual patent grant there could be a few years distance. That's why, in line with relevant literature, we consider priority years for applications and publication years for grants.}.
\end{itemize}

Eventually, the first equation of our IV approach can be written as:

\begin{equation} \label{eq: 1st stage}
    \textit{P}(grant)_{ijht} = \phi_1 \textit{appr}_{jht}^{\ no\ ICT} + \phi_2 \textit{appl}_{jht}^{\ no\ ICT} + \Pi Z_{ijht} + \alpha_h + \alpha_s + \alpha_{jt} + u_{ijht}
\end{equation}

where $\textit{P}(grant)_{ijht}$ is a binary indicator equal to 1 when a firm $i$ at time $t$ is granted property rights by patent office $h$ for a patent in technology class $j$. Besides the instruments mentioned above ($\textit{appr}_{jht}^{\ no ICT}$, $\textit{appl}_{jht}^{\ no ICT}$), we also control for firm-level characteristics $Z_{ijht}$, namely firm size, capital intensity and related growth rates, as well as firm age. $\alpha_h$, $\alpha_s$, $\alpha_{jt}$ are, respectively, a set of countries' patent offices, industry, and technology-year fixed effects. 

Eventually, the outcome equation can be written as follows:

\begin{equation} \label{eq: 2nd stage}
     Y_{ijht} =\alpha_0 + \alpha_1 \hat{P}(grant)_{ijht}+\alpha_2 Z_{ijht}+\alpha_h + \alpha_s + \alpha_{jt} + v_{ijht}
\end{equation}

where $Y_{ijct}$ is the outcome of firm $i$ operating in year $t$, and $\hat{P}(grant)_{ijht}$ is the predicted outcome from eq. \ref{eq: 1st stage}. Please note that, for the scope of our analysis, we perform four separate models reported in the following tables by forwarding both the outcomes, $Y_{ijt+n}$, and the firm-level covariates, $Z_{ijht+n}$, to observe what happens in the years following the grant, $n=(1,2,3,4)$. Eqs. \ref{eq: 1st stage} and \ref{eq: 2nd stage} are a system of simultaneous equations to be run following a classical optimal generalized method of moments (GMM) with robust standard errors. Results\footnote{Results on additional firm-level outcomes are available upon request. We decided to report only significant results on market shares and profitability for shortness' sake.} Main outcomes of interest for smaller firms are reported in Tables \ref{tab: IV ms} and \ref{tab: IV ROCE}. Non-significant results for bigger firms' market shares are reported in Appendix Table \ref{tab: IV bigger firms}

Evidently, the positive impact of patent grants on a firm's market share of smaller firms remains after we check for the unobserved innovative contents of ICT technologies. Magnitudes are not distant from what we already observed in Table \ref{tab: baseline} for smaller firms.

In the second part of Tables \ref{tab: IV ms} and \ref{tab: IV ROCE}, we report the GMM first-stage results of eq. \ref{eq: 1st stage} after a linear probability model (LPM). Tests for overidentification do not reject our instruments and are reported at the bottom of the table. Usefully, we also find that both the approval rates and the total applications in an ICT technology-country-year cell by firms operating in non-ICT markets correlate well with the endogenous binary outcome. In the case of the total number of applications, the coefficient is negative and significantly associated with the odds that a company obtains a patent grant, in line with our intuition that more crowded technological classes are less innovative at the margin.

In line with evidence from previous analyses, we do not find any significant impact on profitability measured by the ROCE index. More in general, we argue that the ensemble of our findings on profitability and market shares relates to existing frameworks that study the peculiar structure of innovative industries, as pertains to the case of the ICT global firms. From this perspective, our findings hint at the existence of endogenous sunk costs \'a la \citet{Sutton1998}. See also \citet{Shaked_Sutton_1983, Shaked_Sutton_1987}. Briefly, even in the presence of an increase in market shares, profit margins may not follow because firms must sustain the high sunk costs in R\&D needed to keep up with innovation and meet customers' demand for novel and differentiated ICT products.

\begin{table}[H] \centering 
  \caption{Market shares and IPR: an IV approach - smaller firms} 
  \label{tab: IV ms} 
     \resizebox{0.99\textwidth}{!}{%
\begin{tabular}{lcccccc}
\\[-1.8ex]\hline 
\hline \\ \\[-1.8ex] 
  \textit{Dependent variable: (log of) Market share} \\ 
\cline{1-1} 
&(t)   &(t+1)   &(t+2)   &(t+3)   &(t+4)  \\ 
\\[-1.8ex]\hline 
\hline \\[-1.8ex] 
$\hat{P}(grant) $&0.26** &0.14   &0.23   &0.057   &0.37 \\
&(0.12)   &(0.13)   &(0.14)   &(0.14)   &(0.19) \\ 
(log of) Firm size&1.07***&1.09***&1.11***&1.11***&1.16*** \\ 
&(0.015)   &(0.015)   &(0.018)   &(0.021)   &(0.027) \\ 
Firm growth&-0.37***&-0.41***&-0.35***&-0.26***&-0.16** \\  
&(0.042)   &(0.057)   &(0.066)   &(0.068)   &(0.072) \\ 
(log of) Capital intensity&0.15***&0.14***&0.16***&0.16***&0.19*** \\ 
&(0.015)   &(0.014)   &(0.014)   &(0.017)   &(0.020) \\ 
Capital intensity growth&-0.013   &-0.12***&-0.063   &-0.028   &-0.087** \\ 
&(0.027)   &(0.031)   &(0.047)   &(0.040)   &(0.035)  \\ 
Firm age&0.011***&0.0058***&0.0056** &0.0030   &-0.0086*** \\ 
&(0.0020)   &(0.0019)   &(0.0024)   &(0.0027)   &(0.0031) \\ 
\hline \\
 \textit{Dependent variable: ${P}(grant): Yes=1$}\\
 \cline{1-1}
&(t)   &(t+1)   &(t+2)   &(t+3)   &(t+4) \\ 
\hline
Approval rate non-ICT firms&0.52***&0.55***&0.47***&0.52***&0.55*** \\ 
&(0.025)   &(0.035)   &(0.040)   &(0.050)   &(0.057) \\ 
Total applications non-ICT firms&-0.030***&-0.032***&-0.040***&-0.044***&-0.040*** \\ 
&(0.0040)   &(0.0047)   &(0.0049)   &(0.0061)   &(0.0078) \\ 
(log of) Firm size&-0.015** &-0.014** &-0.015** &-0.011   &-0.012  \\ 
&(0.0058)   &(0.0061)   &(0.0068)   &(0.0085)   &(0.010) \\ 
Firm growth&0.0043   &0.027   &0.058** &0.037   &-0.087*** \\ 
&(0.012)   &(0.020)   &(0.025)   &(0.035)   &(0.031) \\ 
(log of) Capital intensity&-0.016***&-0.016***&-0.0051   &-0.0094   &-0.018** \\ 
&(0.0054)   &(0.0059)   &(0.0065)   &(0.0073)   &(0.0082) \\ 
Capital intensity growth&-0.011   &0.015   &-0.015   &0.013   &0.018   \\ 
&(0.0093)   &(0.012)   &(0.015)   &(0.019)   &(0.019) \\ 
Firm age&0.00017   &-0.00012   &0.00082   &0.0034** &0.0043** \\ 
&(0.00089)   &(0.00091)   &(0.0011)   &(0.0014)   &(0.0018) \\

N. obs.&4937   &3568   &2610   &1933   &1421  \\ 
R-squared&0.729   &0.750   &0.777   &0.734   &0.748  \\ 
Adjusted R-squared&0.711   &0.727   &0.753   &0.700   &0.712 \\ 
AIC&12412.7   &9051.9   &6312.0   &4737.1   &3303.2  \\ 
\hline
Hansen's J test&11.45   &8.08   &1.12   &2.88   &0.76 \\ 
LM test statistics&402.7   &283.8   &211.5   &183.5   &139.3 \\ 
\hline
 \multicolumn{3}{l}{\textit{} }
\end{tabular}}
\begin{tablenotes}
\footnotesize
\item Note: The table illustrates the impact of a patent grant on firm-level market shares (outcome equation) after an IV approach that controls (first stage) the endogenous role of innovation contents after exogenous variation on assignees in non-ICT industries that obtain patents in the same technologies, countries, and years of ICT firms. At the bottom of the table, we record tests on the validity of instruments. Estimates are obtained after an optimal GMM estimation with robust standard errors in parentheses. $^{*}$p$<$0.1; $^{**}$p$<$0.05; $^{***}$p$<$0.01.
\end{tablenotes}
\end{table}
\newpage

\begin{table}[H] \centering 
  \caption{Profit margins (ROCE) and IPR: an IV approach - smaller firms} 
  \label{tab: IV ROCE} 
     \resizebox{0.99\textwidth}{!}{%
\begin{tabular}{lccccc}
\\[-1.8ex]\hline 
\hline \\ \\[-1.8ex] 
 \textit{Dependent variable: Return on capital employed (ROCE)} \\ 
\cline{1-1} 
&(t)   &(t+1)   &(t+2)   &(t+3)   &(t+4) \\
\\[-1.8ex]\hline 
\hline \\[-1.8ex] 
$\hat{P}(grant) $&-0.008   &0.170   &-0.170   &-0.191   &0.430  \\ 
&(0.21)   &(0.24)   &(0.28)   &(0.28)   &(0.32) \\ 
(log of) Firm size&0.038*  &0.084***&0.10***&-0.037   &0.093* \\
&(0.023)   &(0.026)   &(0.029)   &(0.033)   &(0.052) \\
Firm growth&0.220***& 0.331***& 0.11& 0.222** & 0.240* \\ 
&(0.060)   &(0.085)   &(0.097)   &(0.10)   &(0.13) \\ 
(log of) Capital intensity&-0.151***&-0.140***&-0.120***&-0.111***&-0.099*** \\ 
&(0.020)   &(0.023)   &(0.024)   &(0.027)   &(0.035) \\
Capital intensity growth&0.17***&0.130** &-0.090   &-0.004   &0.077 \\
&(0.045)   &(0.053)   &(0.070)   &(0.077)   &(0.064) \\ 
Firm age&-0.005   &-0.013***&-0.008*  &0.013** &-0.009 \\ 
&(0.003)   &(0.004)   &(0.004)   &(0.006)   &(0.008)  \\ 
\hline \\
 \textit{Dependent variable: $P(grant): Yes = 1 $} \\ 
\cline{1-1}
&(t)   &(t+1)   &(t+2)   &(t+3)   &(t+4) \\
\hline
Approval rate non-ICT firms&0.621***&0.530***&0.511***&0.490***&0.601*** \\
&(0.042)   &(0.047)   &(0.053)   &(0.065)   &(0.076) \\
Total applications non-ICT firms&-0.018***&-0.034***&-0.030***&-0.049***&-0.051*** \\
&(0.006)   &(0.006)   &(0.007)   &(0.008)   &(0.010) \\
(log of) Firm size&-0.017** &-0.010   &-0.020** &-0.012   &-0.023* \\
&(0.007)   &(0.008)   &(0.009)   &(0.011)   &(0.014) \\
Firm growth&-0.005   &-0.010   &0.015   &0.050   &-0.096** \\
&(0.018)   &(0.026)   &(0.034)   &(0.048)   &(0.041)  \\
(log of) Capital intensity&-0.005   &-0.004   &-0.009   &-0.007   &-0.021* \\
&(0.008)   &(0.008)   &(0.009)   &(0.009)   &(0.011)  \\
Capital intensity  growth&-0.026*  &0.001   &-0.024   &0.016   &0.022  \\
&(0.013)   &(0.015)   &(0.019)   &(0.025)   &(0.023)  \\
Firm age &0.001   &-0.001   &0.003   &0.003 \\
&(0.001)   &(0.001)   &(0.001)   &(0.002)   &(0.002)  \\
\hline
N. obs.&2579   &2203   &1746   &1316   & 984 \\
R-squared&0.035   &0.037   &0.042   &0.021   &-0.014 \\
Adjusted R-squared&-0.072   &-0.082   &-0.097   &-0.141   &-0.202  \\
AIC&8006.0   &7071.9   &5353.7   &4032.0   &3035.9 \\
\hline
Hansen's J test&0.34   &2.03   &13.6   &0.18   &0.070 \\
LM test statistic&222.2   &182.9   &129.0   &122.0   &103.8 \\
\hline
 \multicolumn{3}{l}{\textit{} }
\end{tabular}}
\begin{tablenotes}
\footnotesize
\item Note: The table illustrates the impact of a patent grant on firm-level profitability (outcome equation) measured by Returns On Capital Employed (ROCE) after an IV approach that controls (first stage) the endogenous role of innovation contents after exogenous variation on assignees in non-ICT industries that innovate in the same technologies, countries, and years of ICT firms. At the bottom of the table, we report the standard Sargan-Hansen tests and Kleinbergen-Paap LM statistics for overidentification. Estimates are obtained after an optimal GMM estimation with robust standard errors in parentheses. $^{*}$p$<$0.1; $^{**}$p$<$0.05; $^{***}$p$<$0.01.
\end{tablenotes}
\end{table}
\newpage

\section{Robustness and sensitivity checks}\label{sec: robustness}

In this Section, we comment on a few robustness and sensitivity checks with a focus on smaller firms. First, we checked whether the corporate perimeter matters for the magnitude and significance of the impact of patent grants. As we have an industry with global outreach, we cannot exclude that innovation and patenting activity are delegated to subsidiaries in the same country of headquarters or abroad. The main idea is that multinational companies can control important portfolios of patents and manage them through subsidiaries located in many countries. In bigger groups, considerations about fiscal optimization and local knowledge advantages can bring about the location of highly specialized R\&D labs abroad \citep{bosenberg2017rnd, alstadsaeter2018patent, davies2020patent}. For our purpose, we modified our treatment group to consider as also treated those companies whose subsidiaries have been granted a patent in our analysis period. In the second column of Appendix Table \ref{tab: treatment groups}, we visualize the change in the composition of the treatment group. On the other hand, the control group will encompass companies without patents, either at the headquarters or subsidiary levels. We show in Appendix Tables \ref{tab: DiD parent-subsidiaries patents} and \ref{tab: corporate perimeter shares} how our main tenets on market shares are confirmed.

A second concern related to the computation of market shares. We performed a sensitivity analysis to check whether previous results were mediated exclusively by firms' demography since market entry and exit dynamics may correlate with innovation abilities and IPR. Among other things, we know that market shares, firm size, and the number of competitors may evolve over a product life cycle. In other words, it is possible that the dynamics of IPR interact with an endogenous market selection process. Thus, some firms may find it difficult to outlive the market after competitors have been granted the use of fundamental innovations. On the other hand, new entrants may find it convenient to enter the market when a patent grant expires after they have the chance to imitate fundamental products. To be sure that changing barriers to entry are not driving our previous results, we repeat our previous exercises by computing market shares on a balanced sample, i.e., considering only incumbent firms for which we can estimate market shares within the sample, excluding entry and exit dynamics. Results are reported in Appendix Table \ref{tab: entry and exit shares}. The coefficient of interest in the first year after the grant is still positive and significant, with a magnitude slightly lower than in baseline results.

A third concern relates to the geographical composition of the sample. Although we perform our tests on ICT companies located in 39 countries, we find that there is an uneven concentration of companies across countries. The composition of our sample is \textit{prima facie} consistent with what we know about the geographical concentration of the global ICT industry. Countries like the United States, Japan and South Korea host headquarters of important global market players. On the other hand, in section \ref{sec: data}, we showed how countries of the European Union present a relatively lower competitive advantage in the ability to patent innovations. Although the EU hosts an important number of companies in the ICT industry, these companies have a relatively lower propensity to obtain patent grants. Informed by previous evidence, we repeat our exercises on market shares by separating two subsets. In Appendix Table \ref{tab: non-EU shares}, we provide results for companies operating in the United States, Japan and South Korea, whereas the impact on European firms is registered in Appendix Table \ref{tab: EU shares}. Please note how our results are indeed sensitive to geographic location. When we consider only the US, Japan and South Korea, the impact on market shares is bigger and more persistent, up to the third year after the publication of the grant.

\section{Conclusions}
\label{sec: conclusion}

The global ICT industry is a fundamental source of growth in modern economies. Its products and services are purchased by final consumers who want to upgrade and update on the newest technologies; they are also important inputs in the production processes of many other sectors.

Comprehensibly, the sector attracts the attention of both policymakers and scholars from different fields. Most recently, serious doubts have been raised about an excessive market concentration among a few Big-Tech global players. Antitrust authorities in the US and the European Union continue to investigate whether there is evidence of detrimental effects on social welfare. In this context, we may reconnect with the more general debate about the costs and benefits of the present IPR regimes. 

Specifically, after a quasi-experimental setting, our findings point out that smaller companies are the ones that still benefit from patenting. When they do, after the first years of the grant, they obtain a considerable boost in firm size and market share. Competition authorities and policymakers should acknowledge firm heterogeneity when they think about intervention or IPR reform proposals that are intended to reduce IPR protection. Indeed, in our study, we find confirmation that only a few smaller firms apply for patent grants, in line with what was discussed among others by \cite{SMEscoreboard}. The complexities of the modern IPR system and the cost to go to court in case of infringements are more easily withstood by bigger companies, which show to have 89\% of the patents that have been granted in our analysis period.

Nonetheless, by now, we know from previous extensive literature \citep{Comino_et_al_2018} that ICT is an industry where IPR fragmentation is relevant, so much that it becomes increasingly difficult for companies to \textit{’hack their way through the patent thicket’} \citep{Shapiro_2001}. On the other hand, we also know that ICT is a poster child for an industry where R\&D sunk costs are relevant, and companies need to keep a competitive advantage by continuing to invest in innovation \citep{Sutton1991, Sutton1998}. 

Against the present background, we argue that any IPR reform cannot neglect that firms have heterogeneous advantages from patent grants. Smaller firms struggle relatively more to outlive an R\&D-intensive market. At the same time, smaller firms must handle an institutional environment favouring larger firms, which can better navigate intricate IPR regimes. In our view, important avenues of studies could help understand how an optimal IPR policy can respond to the needs of different categories of firm size, providing them with the right incentives to extend investment in innovation efforts.


\setlength\bibsep{0.5pt}
\bibliographystyle{elsarticle-harv}
\bibliography{bibliography.bib}

\pagebreak

\appendix
\section*{Appendix: Tables and Graphs}
\setcounter{table}{0}
\renewcommand{\thetable}{A\arabic{table}}
\setcounter{figure}{0}
\renewcommand{\thefigure}{A\arabic{figure}}

\onehalfspacing

\begin{table}[H]
\caption{Intellectual property rights in ICT industries}
\centering
\resizebox{
\columnwidth}{!}{%
\begin{tabular}{ p{7cm} p{7cm} p{10cm}  }
 \hline \\
Research stream & Contribution(s) &  Issues \\ \\
 \hline  \hline \\ 
ICT surge & |\citet{Hall_2004}, \citet{Danguy_et_al_2014}, \citet{Eberhardt_et_al_2016}, \citet{Branstetter_et_al_2019}, \citet{Venturini_et_al_2022} & The ICT industry is responsible for an increasing share of patents. In particular, software products and intelligent technologies are relevant for implementing Industry 4.0 revolution, and China emerges as a key player.\\ \hline \\
Changing strategies in IPRs & \citet{Cohen_et_al_2000}, \citet{Hall_Ziedonis_2001}, \citet{Cockburn_MacGarvie_2011}, \citet{Graham_et_al_2009}, \citet{Chih_2021}, \citet{Cappelli_et_al_2023} & Patents are increasingly used as firms' strategic tools against potential entrants, in licensing and litigation; relatively less for direct IPR protection.\\ \hline \\
IPR Fragmentation and patent thickets& \citet{Heller_Eisenberg_2000}, \citet{Shapiro_2001}, \citet{Galasso_Schankerman_2010}, \citet{Graevenitz_et_al_2013}, \citet{Noel_Schankerman_2013, Hall_et_al_2020} & Proliferation of patent filings also a consequence of the fragmentation of intellectual property rights, which gives rise to so-called patent thickets, i.e., patents belonging to many companies protecting overlapping technology.\\ \hline \\
Emergence of a market for patents and different institutional arrangements & \citet{Lerner_Tirole_2004}, \citet{Gilbert_2004}, \citet{Swanson_Baumol_2005} \citet{Lerner_Tirole_2008}, \citet{Gans_Stern_2010}, \citet{Arora_Gambardella_2010}, \citet{Baron_et_al_2014}, \citet{Lerner_Tirole_2015}, \citet{Blind_et_al_2017}, \citet{Blind_et_al_2023} & A market for patents is imperfect because of the appropriability problem: different institutional arrangements emerge including Patent Pools, Standard Setting Organizations, Standard-Essential Patents, Fair Reasonable and Nondiscriminatory Licensing (FRAND), Patent Assertion Entities, Patent Intermediaries.\\ \hline
\end{tabular}
}
\begin{tablenotes}
\footnotesize
\item Note: The table reports a few references for understanding the evolution of Intellectual Property Rights in complex technologies, with special attention to the Information and Communication industry.  
\end{tablenotes}
\label{tab: lit review}
\end{table}{}

\begin{table}[H]
\caption{Countries included in the analysis}
\centering
\begin{tabular}{lccccc}
\hline \hline \\
Austria & France & Lithuania & Slovenia \\
Belgium & Germany & Luxembourg & South Korea \\
Brazil & Greece & Malta & Spain \\ 
Bulgaria & Hungary & Netherlands & Sweden \\
Canada &  India & Norway & Switzerland \\
China & Ireland & Poland & Taiwan \\
Croatia & Israel & Portugal & Turkey \\
Czech Republic & Italy & Romania & United Kingdom \\
Denmark & Japan & Russia & United States \\
Finland & Latvia & Slovakia &
 
\\  \hline \hline
\end{tabular}%
\begin{tablenotes}
\footnotesize
\item Note: The table reports the geographic coverage of the firm-level sample for which financial information is available.
\end{tablenotes}

\label{tab: countries}
\end{table}

\begin{table}[H]
\caption{The ICT perimeter based on NACE rev. 2 industries}
\centering
\resizebox{1\columnwidth}{!}{%
\begin{tabular}{lccccc}
\hline \hline \\
NACE Rev. 2 & Description  \\ \hline \\ 
26.1 & Manufacture of electronic components and boards \\
26.2 & Manufacture of computers and peripheral equipment & \textbf{ICT manufacturing}\\
26.3 & Manufacture of communication equipment \\
26.4 & Manufacture of consumer electronics \\ \hline
58.2 & Software publishing  \\
61 & Telecommunications \\
62 & Computer programming, consultancy and related activities & \textbf{ICT services} \\
63.1 & Data processing, hosting and related activities; web portals \\
95.1 & Repair of computers and communication equipment \\
 
\\  \hline \hline
\end{tabular}%
}
\begin{tablenotes}
\footnotesize
\item Note: The table reports the NACE rev. 2 industries included in the sample and belonging to the ICT perimeter.  
\end{tablenotes}
\label{tab: sectors}
\end{table}

\begin{table}[!ht]
    \centering
    \caption{Sample firms' characteristics}
    \resizebox{0.75
\columnwidth}{!}{%
    \begin{tabular}{lccccc}
    \hline
        Variables & ~ & median & mean & st. dev. & max \\ \hline
        Labor productivity & ~ & 79.7 & 220.1 & 7,807.4 & 4,331,715.0 \\ 
        Value added & ~ & 352.7 & 23,215.3 & 705,547.2 & 104,126,578.0 \\ 
        Costs of employees & ~ & 131.8 & 3,650.6 & 68,900.4 & 14,911,167.2 \\ 
        Material costs & ~ & 82.1 & 33,509.3 & 825,637.0 & 132,776,341.5 \\ 
        Fixed assets & ~ & 51.6 & 45,215.2 & 1,781,143.2 & 367,200,100.3 \\ 
        Revenues & ~ & 447.0 & 46,233.0 & 1,249,917.9 & 237,115,000.7 \\ 
        Firm age & ~ & 10.0 & 12.1 & 9.9 & 181.0 \\ 
        Capital intensity & ~ & 8.4 & 308.7 & 29,600.0 & 18423.9 \\ 
        Number of patents & ~ & 0 & 1.96 & 94.55 & 16,708 \\ \hline
    \end{tabular}}
\begin{tablenotes}
\footnotesize
\item Note: The table provides summary statistics of sample firms' main variables that are used in the rest of the paper. Monetary values are expressed in thousand dollars. 
\end{tablenotes}
\end{table}

\begin{table}[H]
\caption{Firm-level outcomes and number of patents. Correlations.}
\centering
\resizebox{0.5
\columnwidth}{!}{%
\begin{tabular}{lccc}
\hline \\
Variable & Coeff. & s. e, &  N. obs. 
\\
\hline \hline \\
Market share & .043*** & (.006) & 931,613 \\ \\
Labor productivity  & .043*** & (.006)  & 931,613  \\ \\
Firm size   & .083***  & (.007) & 931,613      \\    \\
Capital intensity & .068*** & (.006) & 931,613  \\ \\
ROCE (levels) & 2.961 & (3.080) & 884,051 \\ \\
\hline
    \end{tabular}
}
\begin{tablenotes}
\footnotesize
\item Note: Each coefficient results from a least-square regression of the (log) firm-level outcomes (by row) \textit{vis \'a vis} the number of patents granted each year. We scale the number of patents by using the hyperbolic sine transformation ($ln(x+\sqrt{x^2+1})$), which allows approximating to the natural logarithm while retaining the zeros \citep{bellemare2020elasticities}. Firm-level controls, firm-level fixed effects, and country-year fixed effects are included. Standard errors are clustered at the firm level. *** stands for $p < 0.01$.
\end{tablenotes}
\label{tab: OLS parents' patents}
\end{table}{}

\begin{table}[H]
\caption{Treatment group: baseline and robustness check, when considering the corporate perimeter}
\centering
\resizebox{0.5\columnwidth}{!}{
\begin{tabular}{lccccc}
\hline
Year & \multicolumn{1}{p{3cm}}{\centering (1) \\ Treated companies (baseline)} & \multicolumn{1}{p{3cm}}{\centering (2) \\Treated companies (robustness)} \\ \hline \\
2010 & 64 & 65 \\
2011 & 70 & 70 \\
2012 & 52 & 56 \\
2013 & 44 & 46 \\
2014 & 57 & 66 \\
2015 & 49 & 52 \\
2016 & 49 & 54 \\
2017 & 47 & 45 \\
\hline \\
Total & 432 & 454 \\
 
\hline \hline
\end{tabular}
}
\end{table}
\begin{tablenotes}
\footnotesize
\singlespacing
\item Note: Column (1) counts companies granted patents in 2010-2017, i.e., belonging to the treatment group. Column (2) counts companies granted patents at the headquarters or subsidiary level from 2010-2017. Please note that we cannot consider treated companies granted patents in 2009 because we do not have any period before the treatment.
\end{tablenotes}
\label{tab: treatment groups}

\begin{table}[H]
\caption{Firm-level outcomes and patenting activity in the corporate perimeter. Correlations.}
\centering
\resizebox{0.5
\columnwidth}{!}{%
\begin{tabular}{lccc}
\hline \\
Variable & Coeff. & s. e. &  N. obs. 
\\
\hline \hline \\
Market share & .040***  & (.005) & 931,613   
 \\ \\
Labor productivity  &  .040*** & (.005)  & 931,613 \\ \\
Firm size   & .079***  & (.007)  & 931,613   \\    \\
Capital intensity & .070*** & (.006)  & 931,613 \\ \\
ROCE & -.003** & (.001) & 884,051  \\ \\
\hline
    \end{tabular}
}
\begin{tablenotes}
\footnotesize
\item Note: Each coefficient is the result of a least-square regression of the (log) firm-level outcome (by row) \textit{vis \'a vis} the sum of patents granted each year in the company and their subsidiaries standardized with the inverse of an inverse hyperbolic sine transformation ($ln(x+\sqrt{x^2+1})$) to approximate the natural logarithm while retaining zeros, as suggested by \citep{bellemare2020elasticities}. Firm-level controls, firm-level fixed effects, and country-year fixed effects are included. Firm-level clustered standard errors are reported.
\end{tablenotes}
\label{tab:OLS all patents corporate perimeter}
\end{table}{}

\begin{table}[htbp]
\centering
\caption{Balancing properties test after the IPW matching - baseline}
\resizebox{0.8\textwidth}{!}{%
\begin{threeparttable}
\begin{tabular}{lcccccccccc}
\toprule
Treatment time & Covariate &  \multicolumn{2}{c}{Standardized mean difference} & \multicolumn{2}{c}{Variance Ratio} &  & \multicolumn{2}{c}{Effective sample sizes} \\
\cmidrule(lr){3-11}
& & Unadjusted & Adjusted & Unadjusted & Adjusted & \multicolumn{2}{c}{Unadjusted} & \multicolumn{2}{c}{Adjusted} &  \\
\cmidrule(lr){7-11}
& &  &  &  &  & Control & Treated & Control  & Treated &   \\

\midrule
g=2010 & (log) Capital intensity & 0.6332  & 0.0017  &  1.1349 & 1.1703 &  &  & & \\
\addlinespace
& (log) No of employees & 0.8281 &  -0.0042 &  1.5246 & 0.7808 & & & & \\
\addlinespace
& (log) Age & 0.4554 & 0.0102  & 1.1008  & 1.0201 & & & & \\ 
&  &  &   &   & & 216874 & 512 & 22561.92 & 512 \\ 

\midrule
g=2011 & (log) Capital intensity &  0.7408 & -0.0118  & 0.7090  & 0.7162 & & & & \\
\addlinespace
& (log) No of employees & 1.0662 & -0.0136  & 1.5286  & 0.7591 & & & & \\
\addlinespace
& (log) Age & 0.3696 & -0.0036  & 1.1765  & 0.9945 & & & & \\  
&  &  &   &  & & 216950 & 490 & 15458.95 & 490  \\ 
\midrule
g=2012 & (log) Capital intensity &  0.6894 & -0.0059  & 0.8569  & 0.8728 & & & & \\
\addlinespace
& (log) No of employees & 0.7519 &  -0.0089 &  1.5313 & 0.7860 & & & & \\
\addlinespace
& (log) Age & 0.4691 &  0.0022 &  0.6705 & 0.6518 & & & & \\ 
&  &  &   &   & & 216966 & 312 & 26468.02 & 312 \\ 
\midrule
g=2013 & (log) Capital intensity & 1.0022  & -0.0038  & 0.6314  & 0.6021 & & & & \\
\addlinespace
& (log) No of employees & 0.7260 & -0.0015  &  1.5584 & 0.8249 & & & & \\
\addlinespace
& (log) Age & 0.4942 &  0.0024 & 0.6627  & 0.6349 & & & & \\  
&  &  &   &   & & 216986 & 220 & 24566.48 &  220 \\ 
\midrule
g=2014 & (log) Capital intensity &  0.6675 & -0.0019  & 0.5395  & 0.5910 & & & & \\
\addlinespace
& (log) No of employees & 1.0639 & -0.0072  & 0.8737  & 0.4686 & & & & \\
\addlinespace
& (log) Age & 0.8309 & -0.0039  & 0.3716  & 0.3860 & & & & \\ 
&  &  &   &  & & 217095 & 228 & 35632.29 &  228 \\ 
\midrule
g=2015 & (log) Capital intensity &  0.5772 &  -0.0039 & 0.9107  & 0.9896 & & & & \\
\addlinespace
& (log) No of employees & 1.2165 & -0.0136  & 1.3882  & 0.6134 & & & & \\
\addlinespace
& (log) Age & 0.7009 & -0.0031  &  0.5751 & 0.5386 & & & & \\  
&  &  &   &   & & 217104 & 147 & 13797.53 & 147 \\ 
\midrule 
g=2016 & (log) Capital intensity & 0.8063  & 0.0001  & 0.7805  & 0.8188 & & & & \\
\addlinespace
& (log) No of employees & 1.2993 &  -0.0059 & 1.0698  & 0.5206 & & & & \\
\addlinespace
& (log) Age & 0.8952 &  -0.0022 & 0.5965  & 0.5798 & & & & \\ 
&  &  &   &   & & 217153 & 98 & 15466.56 & 98 \\ 
\midrule
g=2017 & (log) Capital intensity &  0.6275 & -0.0015  &  1.0588 & 1.0576 & & & & \\
\addlinespace
& (log) No of employees & 1.2528 &  -0.0043 &  1.0670 & 0.4885 & & & & \\
\addlinespace
& (log) Age & 0.9115 &  -0.0016 &  0.4274 & 0.4107 & & & & \\  
&  &  &   &   & & 217186 & 47 & 17638.61 & 47 \\ 
\bottomrule
\end{tabular}
\begin{tablenotes}
    \footnotesize
    \item Note: The table includes a test for balancing properties after the matching procedure of the treatment vs the control group. Discrete covariates (fixed effects) on sample industries and geography are excluded from the test but not from the sample matching procedure. Given our panel structure, the estimator by \cite{Callaway_Santanna_2021} returns tests for each cohort of treated data, i.e., for each year that a certain number of firms obtained a patent grant, separately. The standardised mean difference is defined as $\frac{\mu_1 - \mu_0}{\sigma_1}$, where $\mu_1$ and $\mu_0$ are the mean values of treated and control observations, respectively, while $\sigma_1$ is the standard deviation of the treated observations.
\end{tablenotes}
\end{threeparttable}%
}
\label{tab:balancing}
\end{table}

\begin{figure}[H]
\caption{Event study on market shares. The case of smaller firms}
\centerline{\includegraphics[scale=0.3]{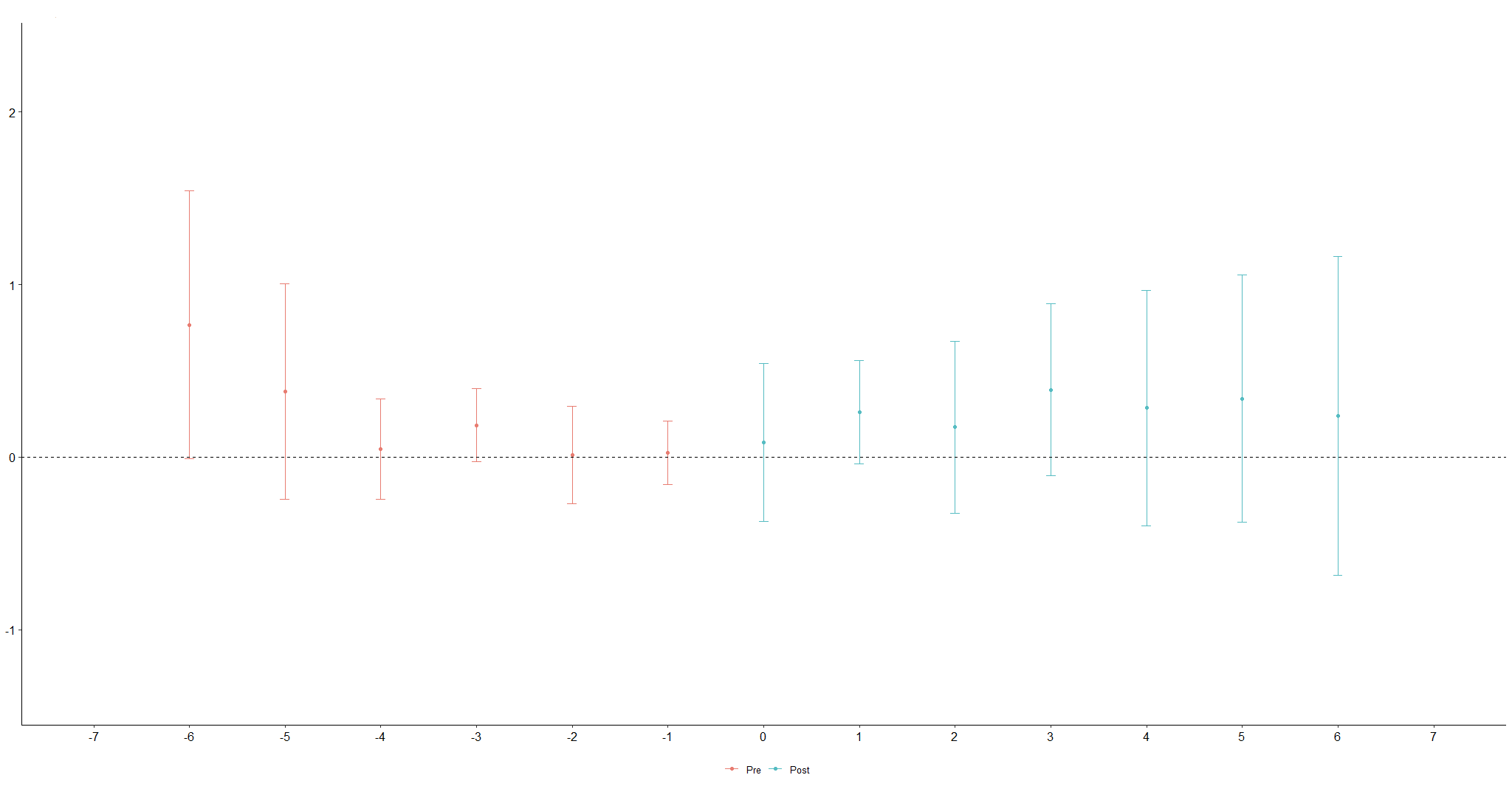}}
\label{fig: MS_ES}
 \begin{tablenotes}
      \footnotesize
      \item Note: Event study on market shares after patent grants following the approach by \citet{Callaway_Santanna_2021}. Smaller firms are below the median at the beginning of the period. Parameters are estimated under parallel trend assumptions conditional on the number of employees, capital intensity, age (in logs), 4-digit sector and regional dummies. Blue lines denote point estimates and simultaneous 99\% confidence bands for the effect of the treatment.
   \end{tablenotes}
\end{figure}

\begin{figure}[H]
\caption{Event study on productivity. The case of smaller firms}
\centerline{\includegraphics[scale=0.5]{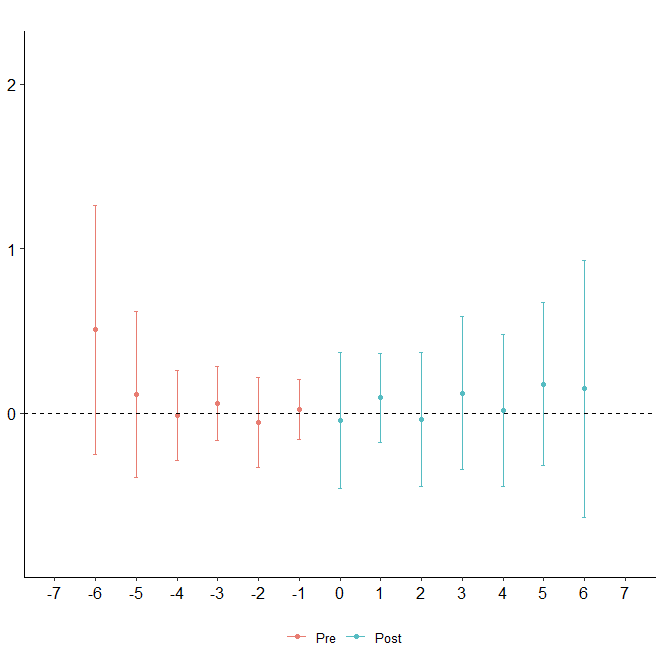}}
\label{fig: LP_ES}
 \begin{tablenotes}
      \footnotesize
      \item Note: Event study on labor productivity measured as value added per employee after patent grants following the approach by \citet{Callaway_Santanna_2021}. Smaller firms are below the median at the beginning of the period. Parameters are estimated under parallel trend assumptions conditional on the number of employees, capital intensity, age (in logs), 4-digit sector and regional dummies. Blue lines denote point estimates and simultaneous 99\% confidence bands for the effect of the treatment.
   \end{tablenotes}
\end{figure}

\begin{figure}[H]
\caption{Event study on firm size. The case of smaller firms}
\centerline{\includegraphics[scale=0.5]{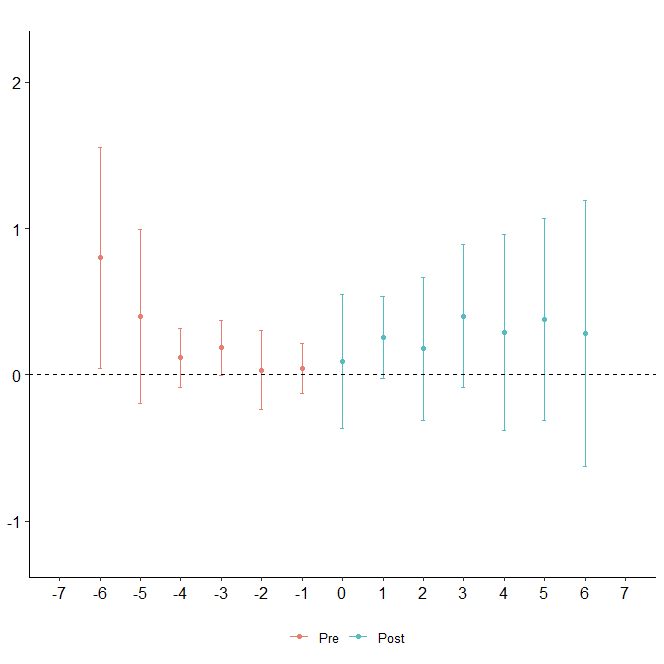}}
\label{fig: Y_ES}
 \begin{tablenotes}
      \footnotesize
      \item Note: Event study on firm size measured as log of revenues after patent grants following the approach by \citet{Callaway_Santanna_2021}. Smaller firms are below the median at the beginning of the period. Parameters are estimated under parallel trend assumptions conditional on the number of employees, capital intensity, age (in logs), 4-digit sector and regional dummies. Blue lines denote point estimates and simultaneous 99\% confidence bands for the effect of the treatment.
   \end{tablenotes}
\end{figure}

\begin{figure}[H]
\caption{Event study on firm size (employees). The case of smaller firms}
\centerline{\includegraphics[scale=0.5]{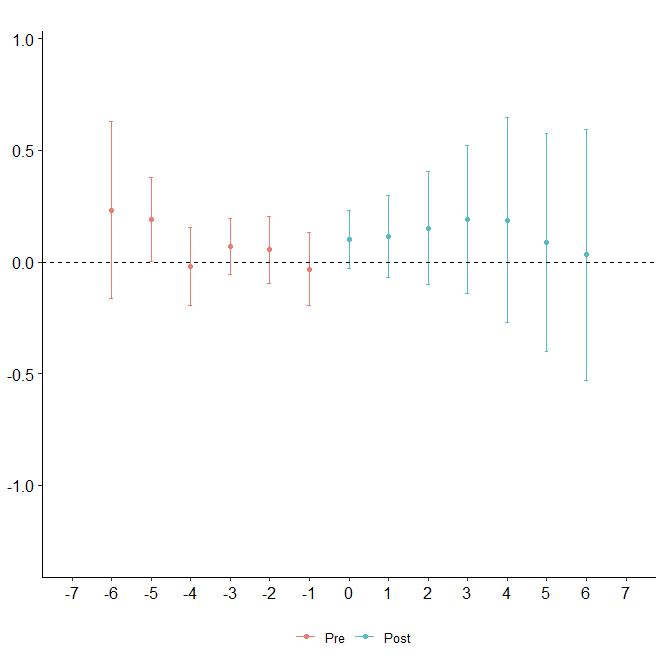}}
\label{fig: L_ES}
 \begin{tablenotes}
      \footnotesize
      \item Note: Event study on the log of the number of employees after patent grants following the approach by \citet{Callaway_Santanna_2021}. Smaller firms are the ones below the median at the beginning of the period. Parameters are estimated under parallel trend assumptions conditional on capital intensity, age (in logs), 4-digit sector and regional dummies. Blue lines denote point estimates and simultaneous 99\% confidence bands for the effect of the treatment.
   \end{tablenotes}
\end{figure}

\begin{figure}[H]
\caption{Event study on capital intensity. The case of smaller firms}
\centerline{\includegraphics[scale=0.5]{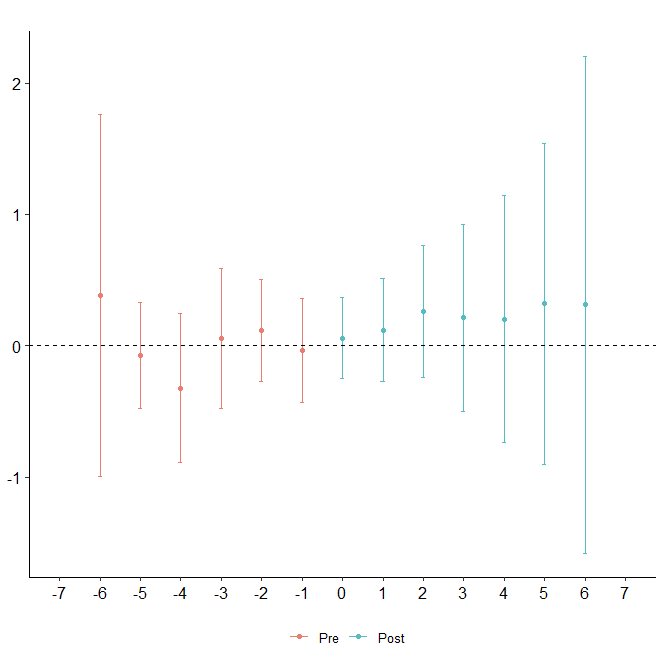}}
\label{fig: KL_ES}
 \begin{tablenotes}
      \footnotesize
      \item Note: Event study on capital intensity after patent grants following the approach by \citet{Callaway_Santanna_2021}. Smaller firms are below the median at the beginning of the period. Parameters are estimated under parallel trend assumptions conditional on the number of employees, age (in logs), 4-digit sector and regional dummies. Blue lines denote point estimates and simultaneous 99\% confidence bands for the effect of the treatment.
   \end{tablenotes}
\end{figure}

\begin{figure}[H]
\caption{Event study on profitability. The case of smaller firms}
\centerline{\includegraphics[scale=0.5]{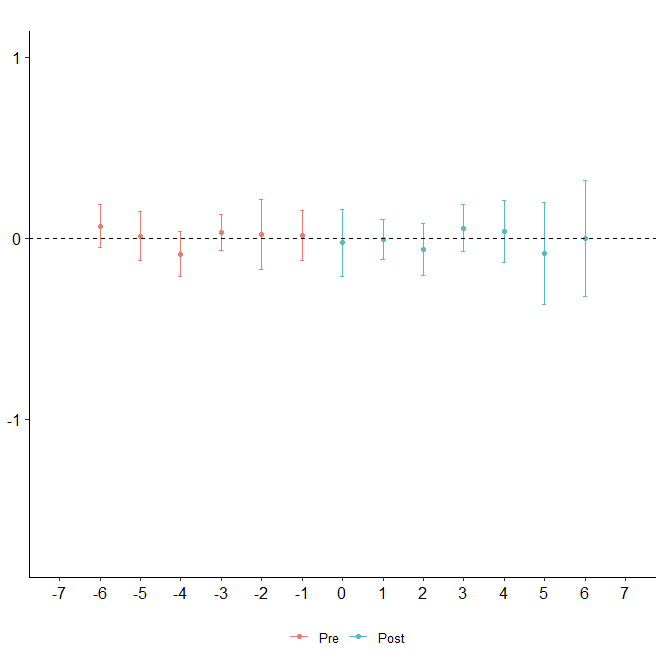}}
\label{fig: ROCE_ES}
 \begin{tablenotes}
      \footnotesize
      \item Note: Event study on profits (ROCE) after patent grants following the approach by \citet{Callaway_Santanna_2021}. Smaller firms are below the median at the beginning of the period. Parameters are estimated under parallel trend assumptions conditional on the number of employees, capital intensity, age (in logs), 4-digit sector and regional dummies. Blue lines denote point estimates and simultaneous 99\% confidence bands for the effect of the treatment.
   \end{tablenotes}
\end{figure}

\begin{figure}[H]
\caption{Variation in the propensity to grant a patent by country patent offices for ICT inventions in 2009-2017}
\centerline{\includegraphics[scale=0.8]{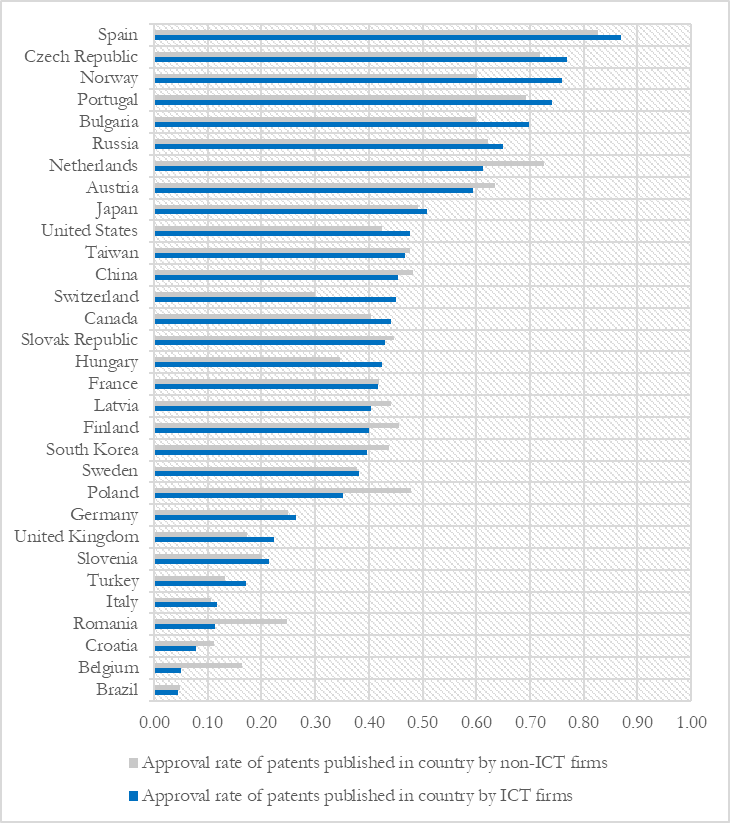}}
\label{fig: IV validity}
 \begin{tablenotes}
      \footnotesize
      \item Note: The figure depicts propensities to grant patents for ICT inventions computed separately for ICT and non-ICT firms at the level of country patent offices. Patent offices with null propensities are excluded. We estimate a $0.827$ correlation between the approval rates of ICT and non-ICT firms.
   \end{tablenotes}
\end{figure}

\newpage

\begin{table}[H]
\caption{Patenting activity and firm-level outcomes, considering the corporate perimeter.}
\centering
\resizebox{0.9
\columnwidth}{!}{%
\begin{tabular}{lcccc}
\hline \\
Variable & Coeff. & s. e. &  No. of treated firms & No. of untreated firms  
\\
\hline \hline \\
(log) Market share & 0.0701** & 0.0286 & 454 & 23,991 \\ \\
(log) Labor productivity  & 0.012 & 0.0205 & 454 & 23,991   \\ \\
(log) Firm size (employees) & 0.0350 & 0.0244 & 454 & 23,991  \\ \\
(log) Firm size (turnover) & 0.0711** & 0.0289 & 454 & 23,991  \\ \\
(log) Capital intensity & 0.0734* & 0.0409 & 454 & 23,991   \\ \\
ROCE (levels) & -0.0205 & 0.014 & 105 & 5,729   \\ \\
\hline
    \end{tabular}
}
\begin{tablenotes}
\footnotesize
\item The table illustrates average aggregate treatment effects after following the doubly robust variant of the method by \citet{Callaway_Santanna_2021}, in the presence of a panel setting, under the assumption of parallel trends conditional on firm-level control variables, 4-digit sector and regional dummies. The treatment group consists of firms with patents granted either at the headquarters or in subsidiaries. Firms are matched using inverse probability weights. Errors are clustered at the firm level. *, ** and *** denotes significance at 10\%, 5\% and 1\% respectively.
\end{tablenotes}
\label{tab: DiD parent-subsidiaries patents}
\end{table}{}

\begin{table}[H] \centering 
  \caption{Market shares and protection of property rights: an IV approach - bigger firms} 
  \label{tab: IV bigger firms} 
     \resizebox{0.99\textwidth}{!}{%
\begin{tabular}{lccccc}
\\[-1.8ex]\hline 
\hline \\ \\[-1.8ex] 
 \textit{Dependent variable: (log of) Market share} \\ 
\cline{1-1} 
&(t)   &(t+1)   &(t+2)   &(t+3)   &(t+4) \\
\\[-1.8ex]\hline 
\hline \\[-1.8ex]
$\hat{P}(grant) $ & 0.010 &	0.051 &	0.161 & 0.111 &	0.097 \\
& (0.014)   &	(0.035)  & 	(0.172)   &	(0.210)   &	(0.179) \\
(log of) Firm size & 0.985***	& 0.999*** &	1.12*** &	1.04*** &	1.12***  \\
&  (0.007)   &	(0.009)   &	(0.010)  & 	(0.011) &  	(0.013)   \\
Firm size growth & -0.234*** &	-0.215*** &	-0.211*** &	-0.157** &	-0.110*** \\ 
& (0.040)   &	(0.056)   &	(0.051)  & 	(0.054) &  	(0.042) \\ 
(log of) Capital intensity & 0.242*** &	0.243*** &	0.251***	& 0.178*** &	0.115*** \\
& (0.010)   &	(0.011)   &	(0.012)   &	(0.015)  & 	(0.019) \\
Capital intensity growth &  -0.036  &	-0.091*** &	0.028 &  	0.012	& 0.010 \\
& (0.035)   &	(0.025)  &	(0.037)   &	(0.031)   &	(0.046) \\
(log of) Firm age &  -0.002*** &	-0.002*** &	-0.004*** &	-0.002*** &	-0.002*** \\
& (0.001) &  	(0.001)  & 	(0.001)   &	(0.001)   &	(0.001) \\
\hline
 \textit{Dependent variable: $P(grant): Yes = 1 $} \\ 
\cline{1-1} 
&(t)   &(t+1)   &(t+2)   &(t+3)   &(t+4) \\
\hline
Approval rate non-ICT & 0.561***	& 0.565*** &	0.562*** & 0.570*** &	0.512*** 	\\
  & (0.031)  & 	(0.037) &  	(0.041)  & 	(0.046)  & 	(0.094)   \\ 
Total applications non-ICT & -0.035*** & -0.032***	& -0.040*** &	-0.040*** & -0.037***\\ 
                                     & (0.002)  & 	(0.006)  & (0.009) &   	(0.006)  &	(0.001) \\
(log of) Firm size   & -0.013***	& -0.009*** &	-0.003 &	-0.002  & 	-0.001   \\
 & (0.002) & 	(0.002) & 	(0.003)  &	(0.0038)  &	(0.0049) \\
Firm size growth &   0.053	& 0.061***	& 0.022***  &	0.016***  & 	0.073\\
 & (0.100)   &	(0.011)  & 	(0.018)  & 	(0.005)  & 	(0.094) \\ 
(log of) Capital intensity & -0.0046  & 	-0.0069** &	-0.0029  & 	-0.0078** &	-0.015*** \\
 &  (0.0030) & 	(0.0033)  &	(0.0034) & 	(0.0038) & 	(0.0047) \\
Capital intensity growth & 0.0046   & 0.023** &	-0.014  & 	0.0100   &	-0.0041    \\
 & (0.0076)  & 	(0.010)  & 	(0.013)  & 	(0.017)  & 	(0.016)  \\
 (log of) Firm age & 0.00020 & 	0.00023  & 	0.00037*  &	0.00033  & 	0.00011   \\
  & (0.00021)   &	(0.00020)   &	(0.00021)   &	(0.00022)  & 	(0.00026) \\
\hline

No. obs, & 4,103  & 	3,905  & 	3,765 &  3,568  & 3,100 \\
R-squared&0.780   &0.800   &0.659   &0.547   &0.711 \\
Adjusted R-squared&0.745   &0.654   &0.539   &0.694   &0.667 \\
AIC&752.9   &412.8   &401.5   &199.9   &178.1  \\
\hline
Hansen J test & 19.5  &	11.2  &	7.11   &	15.45 & 	17.12 \\
LM test statistic &  500.7 &	508.2  &	675.2  & 	537.9 &  	509.4 \\
\hline
 \multicolumn{3}{l}{\textit{} }
\end{tabular}}
\begin{tablenotes}
\footnotesize
\item Note: The table illustrates the impact of a patent grant on firm-level market shares (outcome equation) after an IV approach that controls (first stage) the endogenous role of innovation contents after exogenous variation on assignees in non-ICT industries that innovate in the same technologies and the same year of ICT firms. At the bottom of the table, we report the standard Sargan-Hansen tests and Kleinbergen-Paap LM statistics for overidentification. Estimates are obtained after an optimal GMM estimation with robust standard errors in parentheses. $^{*}$p$<$0.1; $^{**}$p$<$0.05; $^{***}$p$<$0.01.
\end{tablenotes}
\end{table}
\newpage

\begin{table}[H] \centering 
  \caption{Market shares and protection of property rights: an IV approach, considering the corporate perimeter - smaller firms} 
  \label{tab: corporate perimeter shares} 
     \resizebox{0.99\textwidth}{!}{%
\begin{tabular}{lccccc}
\\[-1.8ex]\hline 
\hline \\ \\[-1.8ex] 
 \textit{Dependent variable: (log of) Market share} \\ 
\cline{1-1} 
&(t)   &(t+1)   &(t+2)   &(t+3)   &(t+4) \\
\\[-1.8ex]\hline 
\hline \\[-1.8ex]
$\hat{P}(grant) $ & 0.21** &	0.24 &	0.26 & 0.23 &	0.25 \\
& (0.095)   &	(0.13)  & 	(0.14)   &	(0.15)   &	(0.21) \\
(log of) Firm size & 1.08***	& 1.09*** &	1.10*** &	1.09*** &	1.07***  \\
&  (0.0066)   &	(0.0079)   &	(0.0092)  & 	(0.011) &  	(0.015)   \\
Firm size growth & -0.24*** &	-0.21*** &	-0.23*** &	-0.14** &	-0.10 \\ 
& (0.040)   &	(0.046)   &	(0.046)  & 	(0.063) &  	(0.072) \\ 
(log of) Capital intensity & 0.22*** &	0.23*** &	0.22***	& 0.18*** &	0.15*** \\
& (0.0095)   &	(0.011)   &	(0.012)   &	(0.015)  & 	(0.019) \\
Capital intensity growth &  -0.066*  &	-0.100*** &	0.018 &  	0.12***	& 0.11** \\
& (0.035)   &	(0.025)  &	(0.037)   &	(0.041)   &	(0.046) \\
(log of) Firm age &  -0.0017*** &	-0.0019*** &	-0.0030*** &	-0.0029*** &	-0.0028*** \\
& (0.00045) &  	(0.00047)  & 	(0.00052)   &	(0.00063)   &	(0.00079) \\
\hline
 \textit{Dependent variable: $P(grant): Yes = 1 $} \\ 
\cline{1-1} 
&(t)   &(t+1)   &(t+2)   &(t+3)   &(t+4) \\
\hline
Approval rate non-ICT & 0.60***	& 0.64*** &	0.61*** & 0.60*** &	0.54*** 	\\
  & (0.021)  & 	(0.027) &  	(0.031)  & 	(0.036)  & 	(0.044)   \\ 
Total applications non-ICT & -0.027*** & -0.031***	& -0.039*** &	-0.042*** & -0.039***\\ 
                                     & (0.0032)  & 	(0.0036)  & (0.0039) &   	(0.0046)  &	(0.0060) \\
(log of) Firm size   & -0.013***	& -0.010*** &	-0.0080** &	-0.0022  & 	-0.00031   \\
 & (0.0027) & 	(0.0029) & 	(0.0032)  &	(0.0038)  &	(0.0049) \\
Firm size growth &   0.041***	& 0.059***	& 0.032*  &	0.016  & 	-0.073***\\
 & (0.010)   &	(0.016)  & 	(0.017)  & 	(0.025)  & 	(0.024) \\ 
(log of) Capital intensity & -0.0046  & 	-0.0069** &	-0.0029  & 	-0.0078** &	-0.015*** \\
 &  (0.0030) & 	(0.0033)  &	(0.0034) & 	(0.0038) & 	(0.0047) \\
Capital intensity growth & 0.0046   & 0.023** &	-0.014  & 	0.0100   &	-0.0041    \\
 & (0.0076)  & 	(0.010)  & 	(0.013)  & 	(0.017)  & 	(0.016)  \\
 (log of) Firm age & 0.00020 & 	0.00023  & 	0.00037*  &	0.00033  & 	0.00011   \\
  & (0.00021)   &	(0.00020)   &	(0.00021)   &	(0.00022)  & 	(0.00026) \\
\hline

No. obs, & 8058  & 	6000  & 	4445 &  3258  & 2330 \\
R-squared&0.799   &0.838   &0.839   &0.747   &0.713 \\
Adjusted R-squared&0.745   &0.784   &0.779   &0.624   &0.567 \\
AIC&752.9   &521.9   &393.3   &218.4   &189.3  \\
\hline
Hansen J test & 11.5  &	0.19  &	0.011   &	9.31 & 	5.92 \\
LM test statistic &  610.8 &	428.3  &	324.5  & 	237.9 &  	185.1 \\
\hline
 \multicolumn{3}{l}{\textit{} }
\end{tabular}}
\begin{tablenotes}
\footnotesize
\item Note: The table illustrates the impact of a patent grant on firm-level market shares (outcome equation) after an IV approach that controls (first stage) the endogenous role of innovation contents after exogenous variation on assignees in non-ICT industries that innovate in the same technologies and the same year of ICT firms. At the bottom of the table, we report the standard Sargan-Hansen tests and Kleinbergen-Paap LM statistics for overidentification. Estimates are obtained after an optimal GMM estimation with robust standard errors in parentheses. $^{*}$p$<$0.1; $^{**}$p$<$0.05; $^{***}$p$<$0.01.
\end{tablenotes}
\end{table}
\newpage

\begin{table}[H] \centering 
  \caption{Market shares and protection of intellectual property rights; an IV approach, only incumbent smaller firms} 
  \label{tab: entry and exit shares} 
     \resizebox{0.99\textwidth}{!}{%
\begin{tabular}{lccccc}
\\[-1.8ex]\hline 
\hline \\ \\[-1.8ex] 
 \textit{Dependent variable: (log of) Market share} \\ 
\cline{1-1} 
&(t)   &(t+1)   &(t+2)   &(t+3)   &(t+4) \\
\\[-1.8ex]\hline 
\hline \\[-1.8ex] 
$\hat{P}(grant) $&0.17***   &-0.19   &0.27   &0.091   &0.17 \\
&(0.06)   &(0.20)   &(0.24)   &(0.22)   &(0.21)  \\
(log of) Firm size&1.03***&1.04***&1.06***&1.05***&0.98*** \\
&(0.031)   &(0.039)   &(0.041)   &(0.069)   &(0.076) \\
Firm growth&-0.40***&0.060   &-0.68***&-0.21   &0.041  \\
&(0.15)   &(0.21)   &(0.18)   &(0.16)   &(0.27)  \\
(log of) Capital intensity&0.12***&0.099***&0.0024   &-0.076   &-0.073   \\
&(0.030)   &(0.031)   &(0.036)   &(0.049)   &(0.060) \\
Capital intensity growth&-0.013   &-0.064   &0.095   &-0.0099   &0.0079 \\
&(0.033)   &(0.048)   &(0.063)   &(0.068)   &(0.091)  \\
Firm age&-0.0088*  &-0.0028   &-0.0040   &-0.014***&-0.018*** \\
&(0.0050)   &(0.0049)   &(0.0045)   &(0.0051)   &(0.0061) \\
\hline \\
\textit{Dependent variable: $P(grant): Yes = 1 $} \\ 
\cline{1-1} 
&(t)   &(t+1)   &(t+2)   &(t+3)   &(t+4) \\
\hline
Approval rate non-ICT firms&0.46***&0.40***&0.39***&0.44***&0.51*** \\
&(0.078)   &(0.10)   &(0.12)   &(0.15)   &(0.18)  \\
Total applications non-ICT firms&-0.022** &-0.030***&-0.037***&-0.028   &-0.038** \\
&(0.0093)   &(0.011)   &(0.014)   &(0.018)   &(0.019)  \\
(log of) firm size&-0.011   &-0.0086   &-0.021   &-0.075*  &-0.061  \\
&(0.023)   &(0.026)   &(0.027)   &(0.044)   &(0.054) \\
Firm growth&-0.041   &-0.038   &0.011   &-0.11   &-0.14 \\
&(0.069)   &(0.087)   &(0.084)   &(0.11)   &(0.26)  \\
(log of) Capital intensity&-0.0071   &-0.014   &-0.021   &0.0016   &0.0015  \\
&(0.021)   &(0.024)   &(0.025)   &(0.036)   &(0.038)  \\
Capital intensity growth&-0.042   &-0.037   &0.022   &0.089*  &-0.093 \\
&(0.030)   &(0.032)   &(0.049)   &(0.053)   &(0.061) \\
Firm age&-0.0011   &0.00054   &0.0012   &0.0045   &-0.00086  \\
&(0.0029)   &(0.0033)   &(0.0036)   &(0.0052)   &(0.0050)  \\
\hline

N. obs.& 681   & 521   & 372   & 252   & 199 \\
R-squared&0.799   &0.838   &0.839   &0.747   &0.713 \\
Adjusted R-squared&0.745   &0.784   &0.779   &0.624   &0.567 \\
AIC&752.9   &521.9   &393.3   &218.4   &189.3  \\
\hline
Hansen's J test&2.13   &0.86   &2.01   &0.36   &0.36 \\
LM test statistic&45.5   &30.5   &28.1   &22.4   &19.8  \\
\hline
 \multicolumn{3}{l}{\textit{} }
\end{tabular}}
\begin{tablenotes}
\footnotesize
\item Note: The table illustrates the impact of a patent grant on firm-level market shares (outcome equation) after an IV approach that controls (first stage) the endogenous role of innovation contents after exogenous variation on assignees in non-ICT industries that innovate in the same technologies and the same year of ICT firms. At the bottom of the table, we report the standard Sargan-Hansen tests and Kleinbergen-Paap LM statistics for overidentification. Estimates are obtained after an optimal GMM estimation with robust standard errors in parentheses. $^{*}$p$<$0.1; $^{**}$p$<$0.05; $^{***}$p$<$0.01.
\end{tablenotes}
\end{table}
\newpage

\begin{table}[H] \centering 
  \caption{Market shares and protection of intellectual property rights: an IV approach, considering smaller firms in the US, Japan, and South Korea} 
  \label{tab: non-EU shares} 
     \resizebox{0.99\textwidth}{!}{%
\begin{tabular}{lccccc}
\\[-1.8ex]\hline 
\hline \\ \\[-1.8ex] 
 \textit{Dependent variable: (log of) Market share} \\
\cline{1-1} 
&(t)&(t+1)&(t+2)&(t+3)   &(t+4) \\
\\[-1.8ex]\hline 
\hline \\[-1.8ex] 
$\hat{P}(grant) $&0.60***&0.61***&0.57***&0.56&0.31 \\
&(0.23)   &(0.23)   &(0.16)   &(0.37)   &(0.31) \\
(log of) Firm size&1.14***&1.17***&1.16***&1.12***&1.27*** \\
&(0.027)   &(0.035)   &(0.042)   &(0.041)   &(0.042)  \\
Firm growth&-0.22***&-0.28** &-0.25** &0.024   &0.16 \\
&(0.053)   &(0.11)   &(0.10)   &(0.11)   &(0.14) \\
(log of) Capital intensity&0.20***&0.25***&0.28***&0.27***&0.28*** \\
&(0.021)   &(0.018)   &(0.021)   &(0.025)   &(0.023) \\
Capital intensity growth&0.00063   &-0.040   &-0.023   &0.17*  &0.26** \\
&(0.043)   &(0.061)   &(0.079)   &(0.095)   &(0.10)  \\
Firm age&-0.0016   &-0.0047   &-0.00031   &-0.0062   &-0.024*** \\
&(0.0035)   &(0.0045)   &(0.0056)   &(0.0060)   &(0.0056) \\
\hline \\
 \textit{Dependent variable: $P(grant): Yes = 1 $} \\
\cline{1-1} 
&(t)&(t+1)&(t+2)&(t+3)   &(t+4) \\
\hline
Approval rate non-ICT firms&0.81***&0.77***&0.73***&0.78***&0.91*** \\
&(0.069)   &(0.071)   &(0.079)   &(0.086)   &(0.11) \\
Total applications non-ICT firms&-0.035*   &-0.059***&-0.055** &-0.12***&-0.28*** \\
&(0.017)   &(0.019)   &(0.025)   &(0.046)   &(0.054) \\
(log of) Firm size&-0.017   &-0.017   &-0.037***&-0.0075   &-0.017 \\
&(0.011)   &(0.013)   &(0.014)   &(0.016)   &(0.019) \\
Firm growth&0.030   &0.035   &0.11** &0.088   &-0.084 \\
&(0.028)   &(0.034)   &(0.046)   &(0.062)   &(0.078) \\
(log of) Capital intensity&-0.0094   &-0.0075   &-0.014   &-0.015   &-0.021 \\
&(0.012)   &(0.012)   &(0.012)   &(0.013)   &(0.013) \\
Capital intensity growth&-0.0059   &0.035   &0.0050   &0.018   &0.023  \\
&(0.022)   &(0.027)   &(0.036)   &(0.048)   &(0.051) \\
Firm age&0.0011   &-0.00012   &0.0026   &0.00062   &0.0019 \\
&(0.0018)   &(0.0020)   &(0.0022)   &(0.0027)   &(0.0033)  \\
\hline
N. obs.&1496   &1295   &1046   & 859   & 626 \\
R-squared&0.792   &0.763   &0.746   &0.681   &0.767  \\
Adjusted R-squared&0.768   &0.733   &0.710   &0.633   &0.727  \\
AIC&3534.8   &3203.3   &2602.4   &2173.6   &1393.7 \\
\hline
Hansen's J test&8.82   &4.06   &0.45   &2.56   &0.73 \\
LM test statistic&91.6   &80.0   &61.8   &64.6   &64.6 \\
\hline
 \multicolumn{3}{l}{\textit{} }
\end{tabular}}
\begin{tablenotes}
\footnotesize
\item Note: The table illustrates the impact of a patent grant on firm-level market shares (outcome equation) after an IV approach that controls (first stage) the endogenous role of innovation contents after exogenous variation on assignees in non-ICT industries that innovate in the same technologies and the same year of ICT firms. At the bottom of the table, we report the standard Sargan-Hansen tests and Kleinbergen-Paap LM statistics for overidentification. Estimates are obtained after an optimal GMM estimation with robust standard errors in parentheses. $^{*}$p$<$0.1; $^{**}$p$<$0.05; $^{***}$p$<$0.01.
\end{tablenotes}
\end{table}

\begin{table}[H] \centering 
  \caption{Market shares and protection of intellectual property rights: an IV approach, considering smaller firms in the European Union} 
  \label{tab: EU shares} 
     \resizebox{0.99\textwidth}{!}{%
\begin{tabular}{lccccc}
\\[-1.8ex]\hline 
\hline \\ \\[-1.8ex] 
 \textit{Dependent variable: (log of) Market share} \\
\cline{1-1} 
&(t)&(t+1)&(t+2)&(t+3)   &(t+4)\\
\\[-1.8ex]\hline 
\hline \\[-1.8ex] 
$\hat{P}(grant) $&0.36** &0.21   &0.13   &-0.31   &0.31   \\&(0.14)   &(0.14)   &(0.18)   &(0.21)   &(0.23)   \\(log of) Firm size&1.01***&1.06***&1.11***&1.11***&1.15***\\&(0.023)   &(0.021)   &(0.025)   &(0.036)   &(0.039)   \\Firm growth&-0.42***&-0.49***&-0.47***&-0.59***&-0.29** \\&(0.072)   &(0.093)   &(0.11)   &(0.14)   &(0.12)  \\(log of) Capital intensity&0.12***&0.030   &0.035   &0.047*  &0.084**  \\&(0.023)   &(0.019)   &(0.022)   &(0.026)   &(0.035)   \\Capital intensity growth&-0.036   &-0.18***&-0.042   &-0.040   &-0.15***\\&(0.036)   &(0.046)   &(0.070)   &(0.042)   &(0.036)   \\Firm age&0.026***&0.019***&0.016***&0.011*  &0.00050   \\&(0.0030)   &(0.0031)   &(0.0038)   &(0.0056)   &(0.0056)   \\
\hline
\hline \\ \\[-1.8ex] 
 \textit{Dependent variable: $P(grant): Yes = 1 $} \\
\cline{1-1} 
&(t)&(t+1)&(t+2)&(t+3)   &(t+4)  \\
\hline
Approval rate non-ICT&0.56***&0.51***&0.32***&0.38***&0.44***\\&(0.041)   &(0.049)   &(0.053)   &(0.065)   &(0.080)   \\Total applications non-ICT&-0.031***&-0.037***&-0.051***&-0.047***&-0.039***\\&(0.0051)   &(0.0058)   &(0.0059)   &(0.0069)   &(0.0096)   \\(log of) Firm size&-0.0026   &-0.0085   &-0.0034   &-0.013   &0.00052   \\&(0.0086)   &(0.0089)   &(0.0097)   &(0.012)   &(0.014)   \\Firm growth&-0.022   &-0.069** &0.026   &-0.050   &-0.029   \\&(0.019)   &(0.027)   &(0.040)   &(0.048)   &(0.052)   \\(log of) Capital intensity&-0.012   &-0.013   &0.000012   &0.0016   &-0.0048   \\&(0.0073)   &(0.0082)   &(0.0089)   &(0.010)   &(0.016)   \\Capital intensity growth&-0.021   &0.0031   &-0.021   &-0.020   &0.013   \\&(0.013)   &(0.015)   &(0.020)   &(0.022)   &(0.023)   \\Firm age&0.0019   &0.0037** &0.0022 &0.0074***&0.0094***\\&(0.0014)   &(0.0016)   &(0.0018)   &(0.0021)   &(0.0030)  \\
\hline
N. obs. &2053   &1611   &1179   & 804   & 543     \\R-squared&0.712   &0.793   &0.802   &0.786   &0.812   \\Adjusted R-squared&0.675   &0.761   &0.766   &0.737   &0.761 \\AIC&5354.8   &3709.3   &2665.6   &1725.3   &1053.7 \\
\hline
Hansen's J test&10.5   &15.5   &2.75   &2.62   &0.28   \\LM test statistic &245.6   &183.5   &129.5   &103.1   &64.1 \\
\hline
 \multicolumn{3}{l}{\textit{} }
\end{tabular}}
\begin{tablenotes}
\footnotesize
\item Note: The table illustrates the impact of a patent grant on firm-level market shares (outcome equation) after an IV approach that controls (first stage) the endogenous role of innovation contents after exogenous variation on assignees in non-ICT industries that innovate in the same technologies and the same year of ICT firms. At the bottom of the table, we report the standard Sargan-Hansen tests and Kleinbergen-Paap LM statistics for overidentification. Estimates are obtained after an optimal GMM estimation with robust standard errors in parentheses. $^{*}$p$<$0.1; $^{**}$p$<$0.05; $^{***}$p$<$0.01.
\end{tablenotes}
\end{table}
\newpage

\end{document}